\documentclass[twocolumn,
         showpacs,
         amsmath,
         amssymb,
         amsfonts,
         superscriptaddress,
         aps, 
         prb,
         floatfix,
]{revtex4-2}  
\usepackage[english]{babel}
\usepackage[utf8]{inputenc}   
\usepackage{dsfont}   
\usepackage{SIunits}
\usepackage{hyperref}
\usepackage{graphicx}     
\usepackage{color}
\newcommand{\bigzero}{\mbox{\normalfont\Large\bfseries 0}} 

\begin{document}
\title{Current and shot noise in a normal metal–superconductor junction driven
by spin-dependent periodic pulse sequence}
\author{Bruno Bertin-Johannet}
\email{bruno.bertin@cpt.univ-mrs.fr}
\author{Beno\^it Gr\'emaud}
\author{Flavio Ronetti}
\author{Laurent Raymond}
\author{J\'er\^ome Rech}
\author{Thibaut Jonckheere}
\author{Thierry Martin}
\affiliation{Aix Marseille Univ, Universit\'e de Toulon, CNRS, CPT, IPhU, AMUtech, Marseille, France}

\begin{abstract}
%
  Andreev reflection is a fundamental transport process occurring at the junction between a normal metal and a superconductor (a N-S junction), when an incident electron from the normal side can only be transmitted in the superconductor as a Cooper pair, with the reflection of a hole in the normal metal.  As a consequence of the spin singlet nature of the BCS Cooper pairs, the current due to Andreev reflection at a N-S junction is always symmetric in spin.  Using a Keldysh Nambu Floquet approach, combining analytical and numerical calculations, we study in details the AC transport at a N-S junction, when the two spin components in the normal metal are driven by different periodic drives. We show that, in the Andreev regime, i.e. when the superconducting gap is much larger than the frequency of the drives, the spin-resolved photo-assisted currents are always equal even if the two drives are different. In addition, we show that in this regime the excess noise depends only on the sum of the periodic drives, and we consider in particular the case of Lorentzian pulses (Levitons). We also show how these properties get modified when going beyond the Andreev regime. Finally we give a simple analytical proof of the special properties of the Andreev regime using an exact mapping to a particular N-N junction.
\end{abstract}
 
\maketitle 

\section{Introduction}
\label{Intro}
  Electron quantum optics (EQO) aims at describing and manipulating single electronic excitations in condensed matter systems~\cite{bocquillon14,bauerle18}. 
  This is achieved by adapting scenarios of quantum optics like the Hanbury-Brown and Twiss experiment~\cite{brown1956a} where the intensity correlations from coherent photons at the output are observed or the Hong-Ou Mandel setup,~\cite{hong1987a} where photons collide at the location of the beam splitter and correlations are measured at the output. 
  In condensed matter settings, electron wave guides can be achieved with a two dimensional electron gas, while a quantum point contact mimics the beam splitter. 
  However, electrons differ from photons as they are charged particles and bear fermionic statistics. 
  This means, in particular, that they interact strongly with their neighboring electromagnetic environment and are always accompanied by a Fermi sea.
  In recent decades, the combination of theoretical~\cite{levitov1996a} and experimental~\cite{dubois2013b} efforts, boosted by advances in fabrication techniques, has provided EQO with a strong foothold. 
  Even if EQO has been initially studied in situations when the role of electronic interactions is neglected or minimized~\cite{dubois2013a, dubois2013b}, it is nowadays also studied in strongly correlated systems such as the Fractional Quantum Hall effect~\cite{grenier13,rech2017a} and hybrid superconducting devices~\cite{acciai2019a,Ronetti2020,Bertin2022}.

  In particular EQO has flourished due to the availability of single electron sources working in an AC regime~\cite{lesovik94,kouwenhoven94}, such as the mesoscopic capacitor \cite{feve2007a}, or voltage tailored trains of Lorentzian wave packets  called ``Levitons'',  \cite{levitov1996a,levitov97,dubois13,dubois2013b,jullien14,glattli16,glattli17}.
  These Levitons consist of ``pure'' single electron excitations~\cite{keeling2006a}, i.e., devoid of unwanted electron-hole pairs. 
  For instance, when a combination of AC and DC bias is applied to a device, the measurement of the output excess noise~\cite{keeling2006a,dubois2013b} (with respect to the proper reference situation with only an applied DC bias) allows the detection of these spurious electron-hole excitations.
  Connecting electron waveguides to superconducting leads opened the way to new EQO effects, such as electron (respectively hole) conversion into Bogoliubov quasiparticles~\cite{blonder1982a,beenakker14,ferraro15} above (respectively below) the gap or Andreev reflection~\cite{andreev1964a} (AR) of electrons or holes inside the gap~\cite{acciai2019a}. 

  This normal-superconducting junction was discussed earlier by Belzig \textit{et al.}~\cite{belzig2016a}, where they considered the zero temperature limit and focused on the two limiting regimes where the drive frequency is either much larger or much smaller than the gap of the superconductor. 
  In the latter one, where transport is dominated by Andreev reflection, they found excess noise suppression also for Levitons carrying a half-integer charge. 
  Recently, the intermediate regime has been explored using a microscopic model together with Green's functions in the Keldysh formalism, allowing the computation of the average current as well as the period-averaged noise to all orders in the tunneling constant and at finite temperature~\cite{Bertin2022}. 

  In this article, we extend the preceding setup to a ``conceptual'' situation where spin components are independently driven by periodic sequences of pulses (having the same frequency).
  This additional degree of freedom is like a ``knob'' we can play with, eventually 
  allowing us to  shine a new light on the underlying processes leading to a vanishing 
  excess noise in the Andreev regime.
  More precisely, computing the spin-resolved currents and the excess noise through the normal-superconducting ($N-S$) junction as functions of time, we emphasize that the properties of the junction actually depend only on the total drive.
  For instance, in contrast to a  normal-normal ($N-N$) junction, we find that, independently of the drives imbalance, both spin resolved currents have always the same values,  which is a manifestation of the spin symmetry of the Cooper pairs in the superconductor.  
  This also implies that the total excess noise vanishes as long as the spin-dependent drives actually combine to a proper sequence of Lorentzian pulses, where each pulse injects an integer charge per period. 
  Roughly speaking, it means that the excess noise vanishes as long as the charges injected by each species dependent train of Levitons add to an integer, extending the half-integer charge Levitons situation obtained for balanced drives. 
  A more quantitative analysis of the excess noise as a function of the properties of the two drives for Levitons (injected charges and time delay) allows us to emphasize more precisely the conditions for having a vanishing excess noise.
  
  Even though our main goal is to provide a better understanding of the underlying physics at the $N-S$ junction in the Andreev regime, we would like to mention that, this model could be experimentally achieved by using a quantum spin Hall bar~\cite{Bertin2023} or two half-metals, whose separation needs to be smaller than the superconducting coherence length~\cite{torres1999,deutscher2000}.
  The latter proposition would constitute a new type of Cooper pair beam splitter~\cite{lesovik2001,bouchiat2002,sanchez03,beckmann2004,herrmann10,wei10,das12,baba18}, where typically crossed Andreev reflection~\cite{morten06,morten08} plays a fundamental role.

  The paper is organized as follows. 
  In Sec.~\ref{sec:model} we introduce the theoretical framework for tunneling through the junction in the presence of two periodic pulses with the same period, but driving independently each spin components. 
  In Sec.~\ref{sec:andreev}, we numerically and analytically study the spin-resolved currents and the excess noise in the Andreev regime. 
  We also compute an effective Dyson equation describing the underlying physics as a metal-metal junction, providing an interpretation of our results. 
  We conclude in Sec.~\ref{sec:conclu}. Some additional technical aspects are presented in the Appendices.

  We adopt units in which $\hbar=k_B=1$ and the electronic charge is $e<0$.
  The temperature of the system corresponds to $\Theta$ and $\beta$ denotes the inverse temperature, i.e $\beta^{-1}=k_B\Theta$.
 
\section{Model}
\label{sec:model}
  We adopt a standard approach  developed for junctions involving superconductors
  in which the BdG equations are discretized~\cite{cuevas1999a,Bertin2022}. The metallic and
  superconducting leads are described at equilibrium by tight binding
  Hamiltonians $H_N$ and $H_S$; $H_N$ simply corresponds to the kinetic term, i.e. amounting to the electrons hopping between neighboring sites of a single 1D chain. For the superconducting lead, $H_S$, in addition to the kinetic term, also includes a pairing term, which at the meanfield level reads:
  \begin{equation}
    \Delta\sum_i\left(c_{i \downarrow}^\dagger c_{i\uparrow}^\dagger +c_{i\downarrow}c_{i\uparrow}\right),
  \end{equation}
  where  $i$ labels
  the various sites of these leads, $\Delta$ is the superconducting order parameter and $c_{i\sigma}$  is the electron annihilation operator at a given site $i$ with spin $\sigma$.
  This mean-field approximation is known to provide a very good description of 
  the properties of superconductors in the weak pairing regime which corresponds to the usual situations in mesoscopic physics.

  One defines the Nambu spinors at each boundary of the tunneling junction between the two leads:
  \begin{equation}
    \psi_N^\dagger=\left(c_{N,\uparrow}^\dagger,c_{N,\downarrow}\right)\quad
    \psi_S^\dagger=\left(c_{S,\uparrow}^\dagger, c_{S,\downarrow}\right),
  \end{equation}
  such that the total Hamiltonian reads
  \begin{equation}\label{CurrentHamApp}
    H=H_N+H_S+\psi_N^\dagger W_{NS}\psi_{S}+\text{H.c.}\, ,
  \end{equation}
  where the tunnel matrix from the normal  lead to the superconducting lead reads
  \begin{equation}
  \label{eq:tunnel}
    W_{NS}(t)=\lambda\begin{pmatrix}
      e^{i\phi_\uparrow(t)} & 0\\
      0 & -e^{-i \phi_\downarrow(t)}
    \end{pmatrix}.
  \end{equation}
  $\lambda$ is the tunneling amplitude for both spin species. 
  The functions
  $\phi_{\sigma}(t) = e\int_{-\infty}^{t} dt'\, V_{\sigma}(t')$ are the time-dependent phase differences between the leads accounting for the spin-dependent drives $V_{\sigma}(t)$ applied on the metallic lead. 

\begin{figure}[h]
  \centering
    \includegraphics[width=0.95\columnwidth]{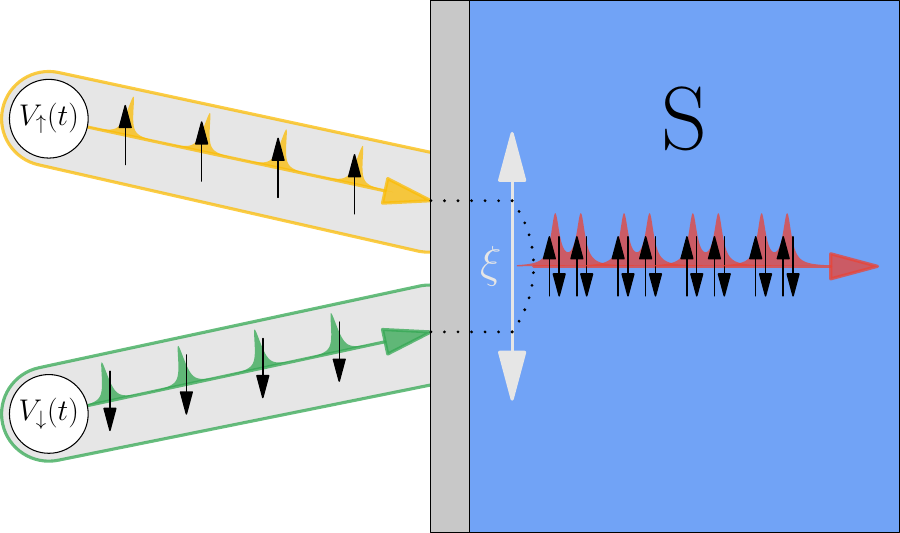}
   \caption{\label{fig:model} (Color online) Model: two half-metal leads are connected to the same superconducting lead. Each half-metal is tuned to allow the propagation of a given spin species only, either 
   $\uparrow$ or $\downarrow$ and is driven by a different voltage source. The superconductor is grounded. The Andreev reflection from one lead to the other is possible as long as the coherence length $\xi$ of the supraconductor is larger than the distance between them. }
  \end{figure}

  This situation could be realized by having a metallic lead actually corresponding to two different leads made of spin-polarized half-metals, allowing us, thereby, to drive each spin component independently. 
  Conceptually, it amounts to having two separate leads for the normal metal side, see Fig.~\ref{fig:model}. 
 Of course, one could have a more realistic description of the tunneling junction, i.e. including the (transverse) geometry of the leads and/or the superconducting order parameter. However, as long as the junction size is much smaller than the superconducting coherence length $\xi$, one can neglect the spatial variation of the superconducting order. Along the same lines, if the transverse size of the junction is small enough (of the order of the Fermi wavelength), i.e. only a few channels are actually coupled through the junction, which amounts to describing the junction by a transmission matrix~\cite{cuevas1999a}. Our tight-binding model provides the simplest parameterization of the transmission coefficient $\tau$ (see below) of the junction. Of course, more elaborate models can be studied. For instance allowing a more complicated tunneling scheme involving more sites would lead to a different relation between the transmission matrix and the microscopic parameters. Still, in the wide band limit, one expects that the microscopic details of the description of the junction should not play an important role as long as the transmission properties are properly taken into account and matching the experimental results.
  
  The  moments of the current operator are computed within the framework of Keldysh theory~\cite{keldysh1965a,rammer86,kamenev11}, defining thereby time ordered Green's functions on the Keldysh contour, such as,
  \begin{equation}
    G^{+-}_{jj'}(t,t')=i\left\langle\psi_{j'}^\dagger(t')\otimes\psi_j(t)\right\rangle,
  \end{equation}
  where $G_{jj'}^{+-}$ is the full Green's function dressed by the tunneling Hamiltonian,  $j$ and $j'$ are lead indices. For instance, the total current flowing in the metallic lead at the junction reads as a Nambu trace:
  \begin{equation}
    \langle I_N(t)\rangle=e\text{Tr}_{\text{N}}\Re\left[\sigma_zW_{NS}(t)G^{+-}_{SN}(t,t)\right]\, ,
  \end{equation}
  where $\Re$ denotes the real part. In addition, one can compute the spin-resolved currents $\langle I_{N\sigma}(t)\rangle$ as follows:
  \begin{equation}
    \langle I_{N\sigma}(t)\rangle=e\text{Tr}_{\text{N}}\Re\left[\frac{(\openone+\sigma\sigma_z)}{2}W_{NS}(t)G^{+-}_{SN}(t,t)\right],
  \end{equation}
  i.e. corresponding to the two diagonal elements of the $2\times2$ matrix 
  $\sigma_z W_{NS}(t)G^{+-}_{SN}(t,t)$. Similar expressions for the real time zero frequency noise correlator can be derived and are given in App.~\ref{app:noisecalc}

  For periodic drives, the voltage drives split in their DC and AC parts:
  \begin{equation}
    V_{\sigma}(t) = V_{\text{DC}\sigma} + V_{\text{AC}\sigma}(t)\, ,
  \end{equation}
  where $V_{\text{DC}\sigma}$ is time-independent, and $V_{\text{AC}\sigma}(t)$ averages to zero over one period
  $T=2\pi / \Omega$ of the periodic drive. The injected charge per spin per period is 
  $q_{\sigma}=eV_{\text{DC}\sigma}/\Omega$. 
  In practice, the DC components of the drives are actually fully taken into account by shifting the Fermi energy of each spin component of the normal metal by $-eV_{\text{DC}\sigma}$, such that one is left dealing only with the AC part of the drives. Using the periodicity of the drives, we introduce the Fourier components 
  $p_{l\sigma}$ of the functions $e^{-i\phi_{\sigma}}$, namely:
  \begin{equation}
    e^{-i\phi_{\sigma}(t)}=\sum_l p_{l\sigma}(q_\sigma)e^{-il\Omega t}.
  \end{equation}
  By doing so, we use Floquet theory~\cite{tien1963a,shirley1965a,pedersen1998a,moskalets2002,rech2017a,Bertin2022} in which
  the total Hamiltonian is separated into an infinite number of independent harmonics in Fourier space. 
  The Floquet theory goes beyond this simple Fourier decomposition. 
  Indeed, it states that as a consequence of the AC drive, the electrons can gain or lose energy quanta leading to the formation of side bands. 
  The Floquet weight $P_{l\sigma}=|p_{l\sigma}|^2$ therefore corresponds to the probability for an incoming electron with spin $\sigma$ to absorb $l$ photons of energy $\Omega$. 
  The voltage biased lead is then better described as a Floquet state, a superposition of Fermi seas, which we now refer to as “Floquet channels”, with shifted chemical potential $\mu_{\sigma}\rightarrow \mu_{\sigma} -eV_{\text{DC}\sigma} + l\Omega$ and an intensity given by the corresponding Floquet weight $P_{l\sigma}$.

  As explained in details in Ref~\cite{Bertin2022}, one can write the Keldysh dressed Green's function as
  \begin{equation}\label{eq:fullGF}
      \begin{aligned}
      G_{SN}^{+-}=&(\mathds{1}-g_{S}^{r}\Sigma_{SN}^{r}g_{N}^{r}\Sigma_{NS}^{r})^{-1} \\
      &\times \big[g_{S}^{+-}+ g_{S}^{r}\Sigma_{SN}^{r}g_{N}^{+-}(\Sigma_{SN}^{a}g_{N}^{a})^{-1}\big]\\
      &\times(\mathds{1}-\Sigma_{SN}^{a}g_{N}^{a}\Sigma_{NS}^{a}g_{S}^{a})^{-1}\Sigma_{SN}^{a}g_{N}^{a},
  \end{aligned}
  \end{equation}
  where, within the Floquet formalism, the different quantities become infinite matrices in the Nambu-harmonic space. For instance, the self-energy $\Sigma^{r,a}$ reads
  \begin{equation}
      \Sigma_{SN,nm}^{r,a}(\omega)=\lambda\begin{pmatrix}
        p_{n-m,\uparrow} & 0\\
        0 & -p_{m-n, \downarrow}^*
      \end{pmatrix}
  \end{equation}
  and, assuming  the large bandwidth limit for the leads, the bare Green's function in the superconductor reads
  \begin{equation}
  g_{S,nm}^{r,a}(\omega)=-\lim_{\delta\to0}\frac{\Delta\sigma_x+(\omega+n\Omega)\openone}{\sqrt{\Delta^2 - (\omega + n\Omega \pm i\delta)^2}}\delta_{nm},
  \end{equation}
  and, in the metal,
  \begin{equation}
  g_{N,nm}^{r,a}=\mp i\openone\delta_{nm}.
  \end{equation}
  The other bare Green's functions, such as $g_{S}^{+-}(\omega)$ and $g_{N}^{+-}(\omega)$ are given in appendix~\ref{app:Dyson}, together with a full expression of the Keldysh dressed Green's function 
  $G_{SN,nm}^{+-}(\omega)$.


\section{Andreev regime}
\label{sec:andreev}
In the Andreev regime $\Delta\gg\Omega$, the bare Green's functions in the superconductor simplify to
\begin{equation}
  g_{S,nm}^{r,a}(\omega)=-\sigma_x\delta_{nm}\text{ and }  \lim\limits_{\Delta\to \infty}g_{S,nm}^{\pm,\mp}(\omega)= 0,
\end{equation}
such that the Keldysh dressed Green's function Eq.~\eqref{eq:fullGF} formally simplifies to
\begin{equation}
\label{eq:fullGF_Andreev2}
    G_{SN}^{+-}=\frac{\lambda^3}{(1+\lambda^4)^2}\left[
    \sigma_x\mathcal{P}T\mathcal{P}^{\dagger}\sigma_x\mathcal{P}
    -\sigma_x \mathcal{P}\mathcal{P}^{\dagger} \sigma_x\mathcal{P}T\right],
\end{equation}
where we have defined $\Sigma=\lambda\mathcal{P}$ and the matrix $T$ describes the thermal distribution of the electrons in the normal lead, see appendix~\ref{app:Dyson}. 
Note, that one could be tempted to simplify the second term using the fact that
$\sigma_x\mathcal{P}\mathcal{P}^{\dagger}\sigma_x=\openone$. However, this expression would then become ill-defined when computing the system properties: since each term taken separately leads to divergent sums, convergence is obtained only after carefully grouping and rearranging terms, which, eventually would lead to unphysical results, more precisely losing the time dependence of the observables, such as the junction currents.

\subsection{Current}

  As explained above the spin-resolved currents are obtained from $W_{NS}(t)G^{+-}_{SN}(t,t)$ which, within the Floquet theory reads:
  \begin{equation}
  \langle I_{\sigma}(t)\rangle=\frac{e}{2\pi}\sum_{n,m}
  \int_{-\frac{\Omega}2}^{+\frac{\Omega}2} d\omega\,e^{-i(n-m)\Omega t} 
  \left(\mathcal{I}^{\sigma}_{nm}(\omega)+\mathcal{I}^{\sigma*}_{mn}(\omega)\right)
  \end{equation}
  with
  \begin{equation}
  \mathcal{I}^{\sigma}_{nm}(\omega)=\sigma\sum_k\left[\Sigma_{NS,nk}(\omega)G^{+-}_{SN,km}(\omega)\right]_{\sigma\sigma}, 
  \end{equation}
  such that, in the Andreev regime, one obtains:
  \begin{equation}
  \begin{aligned}
  \langle I_{\sigma}(t)\rangle=&\frac{e}{2\pi}\frac{\lambda^4}{(1+\lambda^4)^2}\times\\
  &\sum_{n,m}
  \int_{-\frac{\Omega}2}^{+\frac{\Omega}2} d\omega\,e^{-i(n-m)\Omega t} 
  \left(\mathcal{Q}^{\sigma}_{nm}(\omega)+\mathcal{Q}^{\sigma*}_{mn}(\omega)\right)
  \end{aligned}
  \end{equation}
  with
  \begin{equation}
  \mathcal{Q}^{\sigma}_{nm}(\omega)=\sigma\left[
  \mathcal{P}\sigma_x\mathcal{P}^{\dagger}T\mathcal{P}\sigma_x\mathcal{P}^{\dagger}
      -\mathcal{P}\sigma_x\mathcal{P}^{\dagger}\mathcal{P} \sigma_x\mathcal{P}^{\dagger}T
  \right]_{n\sigma,m\sigma}.
  \end{equation}
  Inserting the expressions for $\mathcal{P}$ and $\mathcal{T}$, one obtains that
  \begin{widetext}
  \begin{equation}
    \mathcal{Q}^{\uparrow}_{nm}(\omega)=
    \sum_{r,k,s} p^*_{k-n,\uparrow}p^*_{s-k,\downarrow}p_{r-s,\downarrow}p_{r-m,\uparrow}
    \left[\tanh\left(\frac{\omega+s\Omega+eV_{\text{DC},\downarrow}}{2\Theta}\right)
    -\tanh\left(\frac{\omega+m\Omega-eV_{\text{DC},\uparrow}}{2\Theta}\right)\right]
  \end{equation}
  and a similar expression for $\mathcal{Q}^{\downarrow}_{nm}(\omega)$, see  App.~\ref{app:current}. 
  Note that it is crucial to start from the proper expression~\eqref{eq:fullGF_Andreev2} for the Green's function to finally get the difference between the two $\tanh$ functions, which ensures properly converging sums and integrals. 
  Indeed, performing the following change of variables, $x = \omega + (s+q_{\downarrow})\Omega$ and shifting all the indices by $s$ (except $s$ itself), the sum over $s$ can be carried out, yielding:
  \begin{equation}
  \left\langle I_{\uparrow}(t)\right\rangle = \frac{e}{\pi}\frac{\lambda^4}{(1+\lambda^4)^2}\sum_{n,r,k,m} \int_{-\infty}^{\infty}\mathrm{d}x\, e^{i(m-n)\Omega t}
  p^*_{k-n,\uparrow}p^*_{s-k,\downarrow}p_{r-s,\downarrow}p_{r-m,\uparrow}\left[
  \tanh\left(\frac{x}{2\Theta}\right)-
  \tanh\left(\frac{x+m\Omega-(q_{\uparrow}+q_{\downarrow})}{2\Theta}\right) \right].
  \end{equation}
  The integral can easily be performed and the preceding expression  becomes
  \begin{equation}
  \left\langle I_{\uparrow}(t)\right\rangle = \frac{2e}{\pi}\frac{\lambda^4}{(1+\lambda^4)^2}\sum_{n,r,k,m}
  e^{i(m-n)\Omega t}p^*_{k-n,\uparrow}p^*_{-k,\downarrow}p_{-r,\downarrow}p_{r-m,\uparrow}
  \bigg(q_{\uparrow}+q_{\downarrow}-m\bigg),
  \end{equation}
  which using the definition of the $p_{l\sigma}$ leads to
  \end{widetext}
  \begin{equation}
  \label{eq:currents_andreev}
  \langle I_{\uparrow}(t) \rangle=\frac{e^2}{2\pi}\tau_A
  \left(V_{\uparrow}(t)+V_{\downarrow}(t)\right),
  \end{equation}
  where $\tau_A=4\lambda^4/(1+\lambda^4)^2$ is the so-called Andreev transmission and does not depend on the temperature. In the weak link regime, $\lambda\ll1$, 
  $\tau_A\propto\lambda^4$, i.e. the square of the transmission coefficient of the junction, that one needs to transmit two electrons (one up and one down) to form a full pair in the superconductor.
  Obviously, the expression for $\langle I_{\downarrow}(t) \rangle$ is the same, such that both currents are always equal and proportional to the sum of the applied drives, recovering the fact that the $N-S$ junction, in the Andreev regime, depicts a fully linear behavior. Furthermore, in the Andreev regime, since only processes involving pair creation/annihilation in the superconductors are taking place, it implies that the number of transmitted electrons with different spins must always be equal, and thereby that
  $\langle I_{\uparrow}(t) \rangle=\langle I_{\downarrow}(t) \rangle $. This has to be contrasted with the metallic regime, i.e. $\Delta=0$, where one obtains (see appendix~\ref{app:current} for details)
  \begin{equation}
  \label{eq:currents_normal}
  \langle I_{\sigma}(t) \rangle=2\frac{e^2}{\pi} \tau V_{\sigma}(t),
  \end{equation}
  with the (normal) transmission $\tau=4\lambda^2/(1+\lambda^2)^2$. As expected, each spin current is simply proportional to the respective drive. The fact that one obtains 
  $\tau_A$ from $\tau$ upon replacing $\lambda$ with $\lambda^2$ is simply a consequence that, at every order, the elementary scattering process in the Andreev regime involves two tunneling events through the junction.
  
\begin{figure*}[h!]
    \centering
    \includegraphics[width=0.95\columnwidth]{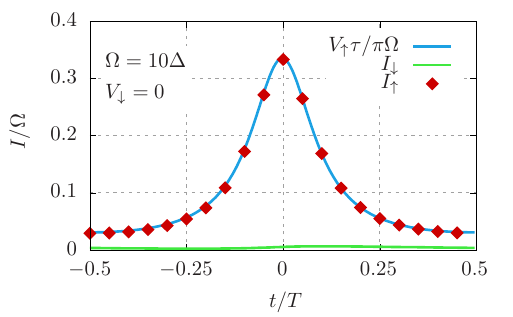}
    \includegraphics[width=0.95\columnwidth]{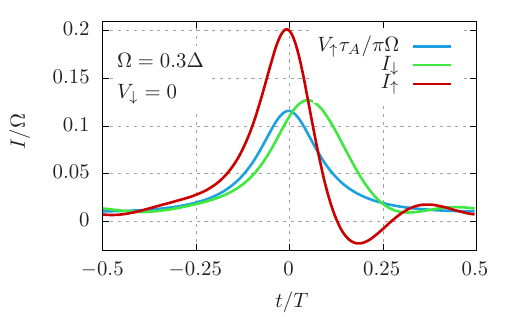}
    \includegraphics[width=0.95\columnwidth]{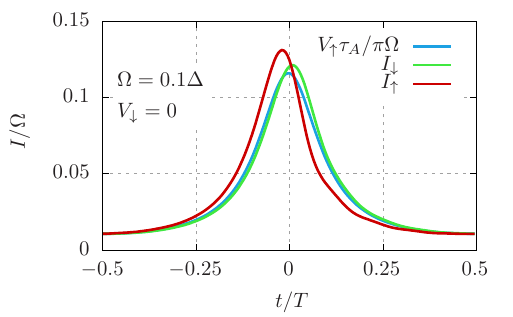}
    \includegraphics[width=0.95\columnwidth]{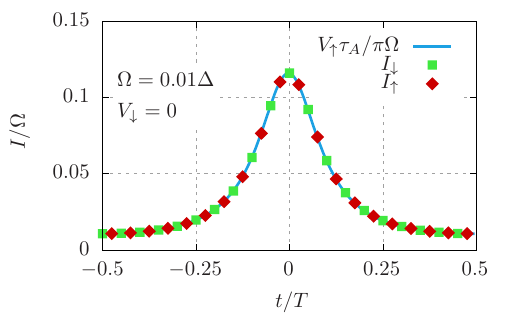}
    \caption{(Color online). Current of spin up and spin down electrons as a function of time for different drive frequencies $\Omega$ where only the up spins are driven, i.e. $V_{\downarrow}=0$. The figure was obtained with the following parameters: $\Theta=10^{-3}$ and $q_\uparrow=1$. The top left plot corresponds to $\Omega=10\Delta$, i.e close to the normal regime. As expected, the spin up current 
    $I_{\uparrow}(t)$ is proportional to $V_{\uparrow}(t)$ where $I_{\downarrow}(t)$ is almost vanishing, of the order of $(\Delta/\Omega)^2$ smaller than $I_{\uparrow}(t)$. On the contrary, in the Andreev regime $\Omega=0.01\Delta$ (bottom right plot), $I_{\uparrow}$ and $I_{\downarrow}$ are equal and proportional to $V_{\uparrow}(t)$. In the intermediate regime $\Omega=0.3\Delta$ (top right plot), one can see the impact of the nonlinear behavior of the junction: both $I_{\uparrow}$ and $I_{\downarrow}$ display a very different behavior as a function of time, with additional oscillations compared to the drive $V_{\uparrow}(t)$.
    Close to the Andreev regime $\Omega=0.1\Delta$ (bottom left plot), both $I_{\uparrow}$ and $I_{\downarrow}$ depart from their Andreev value, but, as expected, the impact of the quasi-particle excitations, directly driven by $V_{\uparrow}$, is stronger on $I_{\uparrow}$.
    \label{fig:time-current}}
  \end{figure*}

  The properties of the currents can be seen in Fig.~\ref{fig:time-current}, which displays $I_{\sigma}(t)$  as a function of time for a vanishing $V_{\downarrow}=0$ and $V_{\uparrow}(t)$ corresponding to a train of Levitons of charge $q_{\uparrow}=1$.  In the Andreev regime $\Omega \ll\Delta$ (bottom right plot), one can see that both currents are  equal and proportional to $V_{\uparrow}(t)$,
  corresponding to Eq.~\eqref{eq:currents_andreev}. Close to the Andreev regime, $\Omega=0.1\Delta$ (bottom left plot), both $I_{\uparrow}$ and $I_{\downarrow}$ depart from their Andreev value, but, as expected, the impact of the quasi-particles excitations, directly driven by $V_{\uparrow}$, is stronger on $I_{\uparrow}$. For increasing values of $\Omega$, reaching the intermediate regime $\Omega\approx\Delta$ (top right plot),  $I_{\uparrow}(t)$ decreases whereas $I_{\downarrow}(t)$ increases, both displaying additional oscillations compared to $V_{\uparrow}$, emphasizing the non-linear behavior of the $N-S$ junction~\cite{Bertin2022}. 
  For $\Omega\gg\Delta$ (top left plot), corresponding to a normal-normal junction,  $I_{\downarrow}(t)$ vanishes 
  (of the order of $(\Delta/\Omega)^2$) , whereas
  $I_{\uparrow}(t)$ becomes again proportional to $V_{\uparrow}(t)$, see Eq.~\eqref{eq:currents_normal}.

\subsection{Excess noise}
\begin{figure}[h!]
  \centering
    \includegraphics[width=0.9\columnwidth]{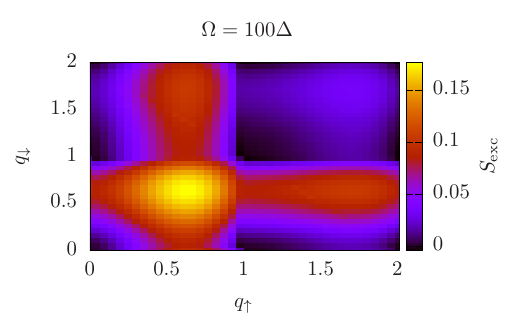}
    \includegraphics[width=0.9\columnwidth]{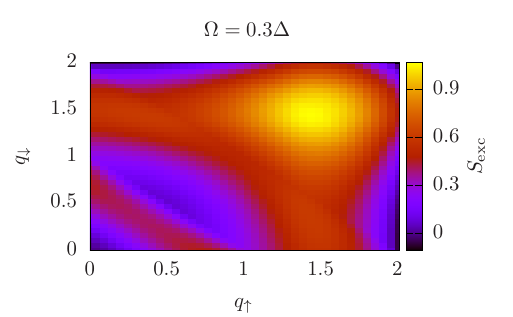}
    \includegraphics[width=0.9\columnwidth]{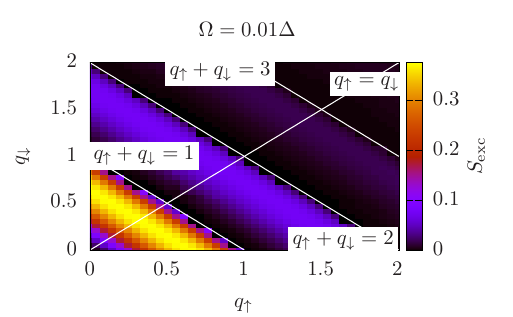}
    
    \caption{(Color online). Excess noise at the junction driven by a Lorentzian drive (of width $\eta=0.1$) as a function of the injected charges $q_\uparrow$ and $q_\downarrow$ and at low temperature ($\Theta=0.001$). The frequency of the drive is, from top to bottom plot, $\Omega=100\Delta$, $\Omega=0.3\Delta$ and $\Omega=0.01\Delta$, corresponding respectively to the normal, intermediate and Andreev regimes. As one can see, in the normal regime, the excess noise only cancels when both Levitons injected charges $q_{\uparrow}=q_{\downarrow}=1$. On the contrary, in the Andreev regime, the excess noise cancels as long as the total injected charge per period
    $q_{\uparrow}+q_{\downarrow}$ is an integer. In the intermediate regime, the excess noise is, in general, non-vanishing.
    \label{fig:xn-lor-imb}}
  \end{figure}

  Finally, along similar lines, one can compute the zero-frequency noise averaged over a period, see Appendix~\ref{app:noisecalc}:
  \begin{widetext}
  \begin{equation}
  \label{eq:noiseandr}
      \overline{\left\langle S\right\rangle}=\frac{e^2}{\pi}
      \left[4\tau^2_A\Theta+
      2\tau_A(1-\tau_A)\sum_{s}(s+q_\uparrow+q_\downarrow)\Omega
      |p^A_s|^2
      \coth\left(\frac{\Omega(s+q_\uparrow+q_\downarrow)}{2\Theta}\right)\right]\, ,
  \end{equation}
  \end{widetext}
  where $p^A_s=\sum_n p_{r,\downarrow}p_{s-r,\uparrow}$. Using the definition of the coefficients $p_{l\sigma}$, we obtain 
  \begin{equation}
    p^A_s=\frac{1}{T}\int_{T/2}^{T/2}\mathrm{d}t\,
      e^{-is\Omega t}e^{i\left[\phi_\uparrow(t)+\phi_\downarrow(t)\right]},
  \end{equation}
  which is therefore the Fourier component of $e^{-i\phi_{\text{tot}}(t)}$, i.e. of the 
  AC part of the \textit{total} drive $V_{\text{AC}\uparrow}(t)+V_{\text{AC}\downarrow}(t)$.
  Therefore, the formula above for $\overline{\left\langle S\right\rangle}$ is the same as the one for  
  zero-frequency noise averaged over a period for an effective normal-normal junction,
  \begin{align}
      \overline{\left\langle S^N\right\rangle}_{q}= & \frac{e^2}{\pi} \bigg[4\tau^2\Theta  +2\tau(1-\tau)\nonumber \\
        \times\sum_n &(eV_{\text{DC}}+n\Omega)|p_n|^2\coth \left(\frac{eV_{\text{DC}}+n\Omega}{2\Theta} \right)\bigg] ,
  \label{eq:normal_noise}
  \end{align}
  driven by $V(t)=V_{\uparrow}(t)+V_{\downarrow}(t)$, i.e. the total drive applied to the NS junction. 
  This emphasizes that, not only for the current, but also for the noise, the behavior of the junction only depends on the total drive. 
  This has an important consequence for the excess noise, 
  in particular when each drive corresponds to a train of Levitons, i.e. a sequence of Lorentzian pulses defined as follows
  \begin{equation}
    V_{\sigma}(t) = V^{\sigma}_{0} \left( \frac{1}{\pi} \sum_k \frac{\eta}{\eta^2 + (t/T - k)^2 }  \right)\, .
  \end{equation}
  where $\eta=W/T$ is the ratio between the width of the pulse $W$ and the period of the drive $T$. In that particular situation, the total drive $V(t)$ simply corresponds to a sequence of Levitons fully characterized by $V_0^{\uparrow}+V_0^{\downarrow}$, i.e. its total charge $q=q_{\uparrow}+q_{\downarrow}$, resulting therefore in a vanishing excess noise when $q$ is an integer. This is a well known results when both drives are the same and each corresponding to a half-Leviton, i.e. 
  $q_{\uparrow}=q_{\downarrow}$, but our computation shows that it extends to any situations where 
  $q_{\uparrow}+q_{\downarrow}$ is an integer. 
  This is exemplified in Fig.~\ref{fig:xn-lor-imb}, where one plots the excess noise, defined as
  $ S_\text{exc} = \overline{\langle S\rangle}  - \overline{\langle S\rangle}_{dc}$, in the 
  $\left(q_{\uparrow},q_{\downarrow}\right)$ plane, for the $N-S$ junction driven by periodic Lorentzian drives, $q_{\sigma}$ corresponding to the injected charge for each spin component per period. 
  $\overline{\langle S\rangle}_{dc}$ is the $DC$ noise computed for the same injected charge.
  As predicted by Eq.~\eqref{eq:noiseandr}, in the Andreev limit $\Omega\ll\Delta$, the excess noise vanishes along lines corresponding to 
  $q_{\uparrow}+q_{\downarrow}\in\mathbb{Z}$. As explained in~\cite{Bertin2022}, for intermediate regimes, the excess noise does not vanish anymore, and, for $\Omega \gg\Delta$, i.e. in the normal-normal regime of the junction, the excess noise only vanishes when both $q_{\sigma}$ are integers. 

\begin{figure*}[h!]
  \centering
    \includegraphics[width=\textwidth]{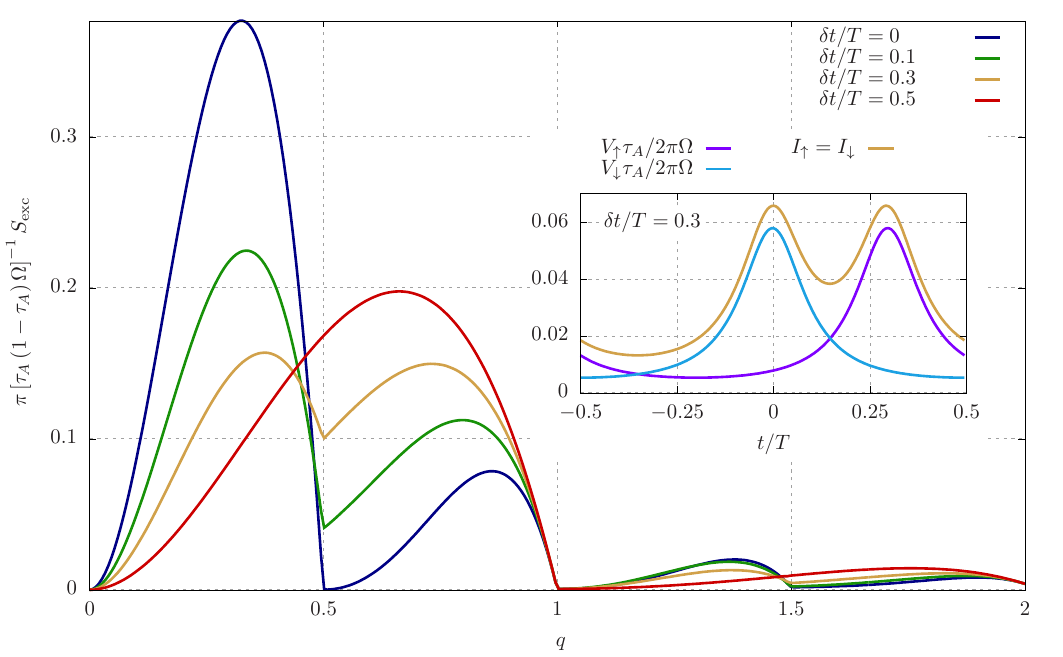}
    \caption{(Color online). Excess noise at the junction driven by a Lorentzian drive (of width $\eta=0.1$) as a function of the injected charge $q=q_{\uparrow}=q_{\downarrow}$ for various time shifts $\delta t$. The frequency of the drive is $\Omega=0.01$ and the reduced temperature is $\Theta=0.001$.
    For integer injected charge $q$, the excess noise always vanishes independently of the time shift $\delta t$, emphasizing that in this case, the \textit{total} drive $V_{\uparrow}+V_{\downarrow}$ corresponds to a sequence of integer Lorentzian pulses. On the contrary, 
    the excess noise for $q=0.5$ is finite as soon as the time shift is non-zero, emphasizing that, in that case, the \textit{total} drive is not a sequence of integer Lorentzian pulse. Only when $\delta t=0$, both pulse sequences $V_{\uparrow}$ and $V_{\downarrow}$  add to a sequence of integer Levitons, resulting in a vanishing excess noise. The inset displays the currents
    $I_{\uparrow}$ and $ I_{\downarrow}$ and the drives  $V_{\uparrow}$ and $V_{\downarrow}$ as functions of time, emphasizing that, in the Andreev regime, the currents are equal and simply proportional to the \textit{total} drive.
    \label{fig:xn-lor-shift}
    }
  \end{figure*}

  In addition, we would like to emphasize that, 
  for the excess noise to vanish in the Andreev regime, our calculations show that the full shape of the total drive $V(t)$ has to correspond to a sequence of Levitons with an integer charge, which is actually a stronger constraint than just having an integer total injected charge.  In the preceding example, where each drive corresponds to Lorentzian pulses centered at the same times $t_k=kT$, this condition was fulfilled as soon as $q_{\uparrow}+q_{\downarrow}$ is an integer. On the other hand, if we consider, for instance, the situation where each drive corresponds to the same sequence of Levitons with a (fractional) charge $q_\sigma$, but being shifted in time one with respect to the other, i.e. 
  $V_{\downarrow}(t)=V_{\uparrow}(t+\delta t)$, then, the total drive $V(t)$  is simply a periodic sequence made of two Levitons per period, each one having a charge $q_\sigma$, which, unless $\delta t=0$, results in a finite excess noise.  This is emphasized in Fig.~\ref{fig:xn-lor-shift}, where the excess noise is plotted as function of $q_\sigma$ for different value of $\delta t$. As one can readily see, 
   for half-integer $q_\sigma$, the excess noise vanishes only when $\delta t=0$. For integer $q_\sigma$, since each drive $V_{\sigma}$ alone will result in a vanishing excess noise independently from the other drive, the total excess noise vanishes for all values of the delay $\delta t$ .
  


  From a physics point of view, it emphasizes the difference between half-integer drives and integer ones: For half-integer drives, i.e. $q_{\sigma}=1/2$ for instance, the Andreev reflection of each spin component produces a ``half-pair'' in the superconductor; each of these ``half-pair'' must then be produced ``at the same time'' to allow for a whole pair to be transmitted in the superconductor. This reasoning goes beyond the half-integer case, and generalizes to any combination $(q_\uparrow, q_\downarrow)$ satisfying $q_\uparrow + q_\downarrow = 1$~\cite{Bertin2023}. On the other hand, for integer drives, the Andreev reflection of each spin component produces a whole pair, independently of the Andreev reflection of the other spin component, allowing for a noiseless current in the junction.

\subsection{Equivalence with an effective metal-metal junction}

The fact that both the current and the noise only depends on the sum of the drives can be directly inferred  from the Dyson equation in the time domain, which reads formally
\begin{equation}
 G(t,t')=g(t-t')+\iint\, dt_1 dt_2 g(t-t_1)W(t_1,t_2)G(t_2.t'),
\end{equation}
where $g$ is the bare Green's function. Every Green's function has the following block structure
\begin{equation}
 G=\left[\begin{array}{c|c}
  G_{NN}^{\eta,\eta'} & G_{NS}^{\eta,\eta'} \\
  \hline
  G_{SN}^{\eta,\eta'} & G_{SS}^{\eta,\eta'}
 \end{array}\right].
\end{equation}
each $G_{ij}^{\eta,\eta'}$ is a $4\times4$ matrix corresponding to Nambu plus Keldysh dimension. The matrix $W$ reads
\begin{equation}
 W(t_1,t_2)=\delta(t_1-t_2)\left[\begin{array}{c|c}
	\bigzero & \begin{array}{c|c} 
	W_{NS} & 0 \\ \hline 0 & -W_{NS}\end{array} \\
	\hline \begin{array}{c|c} 
	W_{SN} & 0 \\ \hline 0 & -W_{SN}\end{array} & \bigzero
                                 \end{array}\right],
\end{equation}
where $W_{NS}$ is the $2\times2$ tunneling amplitude matrix in Nambu space, see Eq.~\eqref{eq:tunnel}.

Iterating once the Dyson equation, one obtains:
\begin{equation}
\begin{aligned}
 G(t,t')&=g(t-t')+g(t-t_1)W(t_1,t_2)g(t_2-t')\\
 &+g(t-t_1)W(t_1,t_2)g(t_2-t_3)W(t_3,t_4)G(t_4,t'),
 \end{aligned}
\end{equation}
where integration over the intermediate times is implicit.

In the Andreev regime, the superconductor bare Green's function simply reads:
\begin{equation}
g^{\epsilon,\epsilon'}_{S}(\tau)=\delta(\tau)\left[\begin{array}{c|c}
      -\sigma_x & 0 \\ \hline 0 & \phantom{0}\sigma_x                          
\end{array}\right],
\end{equation}
where each $2\times2$ sub-blocks are in Nambu space and combined together in the Keldysh space.
Thereby, one obtains the following effective Dyson equation for the normal metal dressed Green's function 
$G_{NN}$:
\begin{equation}
 G^{\epsilon,\epsilon'}_{NN}(t,t')=g^{\epsilon,\epsilon'}_{N}(t-t')+g^{\epsilon,\epsilon'}_{N}(t-t_1)\tilde{W}(t_1)G^{\epsilon,\epsilon'}_{NN}(t_1,t'),
\end{equation}
where 
\begin{equation}
\begin{aligned}
 \tilde{W}(t)=&\left[\begin{array}{c|c} 
	W_{NS} & 0 \\ \hline 0 & -W_{NS}\end{array}\right]
	\left[\begin{array}{c|c}
      -\sigma_x & 0 \\ \hline 0 & \phantom{0}\sigma_x   
\end{array}\right]\left[\begin{array}{c|c} 
	W_{SN} & 0 \\ \hline 0 & -W_{SN}\end{array}\right]\\
	=& \left[\begin{array}{c|c} 
	\tilde{W}_{NN} & 0 \\ \hline 0 & -\tilde{W}_{NN}\end{array}\right],
\end{aligned}
\end{equation}
with  
\begin{equation}
  \begin{aligned}
    \tilde{W}_{NN}(t)&=W_{NS}\sigma_xW_{SN}\\
    &=-\lambda^2
    \left[\begin{array}{c|c}
      0 & e^{i\left(\phi_{\uparrow}(t)+\phi_{\downarrow}(t)\right)} \\ \hline 
      e^{-i\left(\phi_{\uparrow}(t)+\phi_{\downarrow}(t)\right)} & 0                          
    \end{array}\right].
  \end{aligned}
\end{equation}
As one can see, the effective Dyson equation describes a metal-metal junction where one metal is made of spin up and the other of spin down, the effective drive being the sum of the original drives. More precisely, because of the Nambu description of the system, the up spins are electrons, where the down spins correspond to holes. From that point of view, if the sum of the effective drive consist of a train of Levitons of total charge $q=q_{\uparrow}+q_{\downarrow}$, it amounts to converting $q$ electronic charge with spin up to $q$ positive charges with spin down, i.e. removing $q$ electronic charges with down spin as well, corresponding, as expected, to having created $q$ pairs in the superconductor, for a total charge transfer through the junction equal to $2q$. Similar results have been obtained recently when computing the final state in $N-S$ junction driven, in the Andreev regime, by a single Lorentzian pulse~\cite{Bertin2023}. 

Finally, we would like to mention that the present theory predicts that applying opposite drives, i.e. such that
$V_{\uparrow}(t)=-V_{\downarrow}(t)$ results, in the Andreev regime, to vanishing quantities such as currents, excess noise... Therefore, in this configuration, one could have a direct and precise probe of the impact of the finite superconducting gap on the $N-S$ junction properties, close to the Andreev regime.

\section{Conclusion}
\label{sec:conclu}

We have shown from both a numerical and an analytical point of view that, in the Andreev regime, a $N-S$ junction behaves as a normal metal driven by the sum of the drives applied to the junction. More precisely, we have shown that the spin up and spin down currents have always the same value, proportional to the total drive $V_{\text{tot}}$. Similarly, the excess noise only depends on $V_{\text{tot}}$ and vanishes as long as the total drive amounts to a Lorentzian train of pulses injecting an integer number of charges per period. These results are simply explained by mapping the $N-S$ junction to an effective $N-N$ junction driven by $V_{\text{tot}}$. 
The physical origin of this behavior can be traced back to the spin symmetry of the Cooper pairs in the superconductor. In the Andreev regime, they are the only excitations available for the transport, which enforces an equal amount of spins up and spins down flowing through the $N-S$ junction, even if different spin-dependent drives were applied. The present study
mainly focused on the Andreev regime, but it would be worth to have a more thorough analysis of the properties of the system in the non-linear regime $\Omega\simeq\Delta$, i.e. understanding how the fact the two spin species are independently driven impacts the currents and excess-noise.

In addition, we would like to stress  that a possible experimental realization of this system could be achieved using spin polarized half-metals for the normal leads~\cite{Bertin2023}, in particular allowing us to study the impact of the time delay between the drives. 
From that point of view, our microscopic model can be adapted  for a more detailed comparison with experimental results.

Furthermore, these transport properties could be studied using cold atomic gases trapped in optical lattices. In these systems, one could prepare, for instance, an initial wavepacket made of a given species and measure the time-evolution at the boundary between the normal side and the (strongly) paired superconducting side, studying thereby the time-resolved Andreev reflection~\cite{PhysRevLett.100.110404}. Moreover, going beyond the standard fermionic case, Andreev-like reflection can be achieved using Bogoliubov excitations on top of a Bose-Einstein condensate~\cite{PhysRevLett.102.180405}.  
 
Finally, it would be interesting to extend our approach to the case of different types of superconductors (P-wave, topological~\cite{BrunoTS}, high Tc...), where both singlet and triplet pairings must be taken into account~\cite{Prada20}. In this situation, the physics of Andreev reflection is much richer, giving rise to more complex metals involved in the effective description of the junction.

\begin{acknowledgments}


This work received support from the French government under the France 2030 investment plan, as part of the Initiative d'Excellence d'Aix-Marseille Universit\'e A*MIDEX. We acknowledge support from the institutes IPhU (AMX-19-IET008) and AMUtech (AMX-19-IET-01X).
\end{acknowledgments}

\onecolumngrid
\begin{appendix}
\section{Dyson's equation for Keldysh Green's functions.}
\label{app:Dyson}
In this section, we summarize the results derived in~\cite{Bertin2022}.

\subsection{Definitions}
Defining
\begin{equation}
  \begin{aligned}
    \omega_n^{\pm}=\lim_{\delta\to0}\frac{\omega+n\Omega}{\sqrt{\Delta^2 - (\omega + n\Omega \pm i\delta)^2}}\qquad&\qquad\Delta_n^{\pm}=\lim_{\delta\to0}\frac{\Delta}{\sqrt{\Delta^2 - (\omega + n\Omega \pm i\delta)^2}}\\
    \xi_{n}^{\pm}=\frac{1}{1-\lambda^4\left((\omega_n^{\pm})^2-(\Delta_n^{\pm})^2\right)\mp2i\lambda^2\omega_n^{\pm}}\qquad&\qquad\eta_n^{\pm}=\omega_n^{\pm}\mp i\lambda^2\left((\omega_n^{\pm})^2-(\Delta_n^{\pm})^2\right)\\
    \overline{\Delta}_n = \frac{\Delta_n^--\Delta_n^+}{2} \qquad&\qquad \overline{\omega}_n= \frac{\omega_n^- - \omega_n^+}{2}\\
	  \hat{g}_{S,nm}^{r,a}=-\left(\Delta_n^\pm\hat{\sigma}_x+\omega_n^\pm\hat{\mathds{1}}\right)\delta_{nm}\qquad&\qquad\hat{g}_{N,nm}^{r,a}=\mp i\hat{\mathds{1}}\delta_{nm}\\
    \hat{g}_{S,nm}^{\pm\mp}=\left(\overline{\omega}_n\hat{\mathds{1}}+\overline{\Delta}_n\hat{\sigma}_x\right)\left(\tanh(\omega+n\Omega)\mp1\right)\delta_{nm}\qquad&\qquad\hat{g}_{N,nm}^{\pm\mp}=-i\left(\hat{T}\mp\hat{\mathds{1}}\right)\delta_{nm}\\
    \hat{\Sigma}_{SN,nm}^{r,a}=\lambda\begin{pmatrix}
      p_{n-m,\uparrow} & 0\\
      0 & -p_{m-n, \downarrow}^*
    \end{pmatrix}\qquad&\qquad\hat{T}_{nm}=\begin{pmatrix}
	  \tanh\left(\frac{\omega+n\Omega-eV_{\text{DC}\uparrow}}{2\Theta}\right) & 0\\
	    0 & \tanh\left(\frac{\omega + n\Omega+eV_{\text{DC}\downarrow}}{2\Theta}\right)
\end{pmatrix}\delta_{nm}\, .
  \end{aligned}
\end{equation}
the Keldysh dressed Green's function reads:
\begin{equation}
\label{eq:KeldyshGF}
  \begin{aligned}
	  G_{SN,nm}^{\pm\mp}=i\lambda\xi_n^+\xi_r^-\Big[&\sigma_x\mathcal{P}_{nk}  
	  T_k\mathcal{P}_{kr}^\dagger  \mathcal{P}_{rm}  
	  \left[\Delta_n^++i\lambda^2\Delta_n^+\omega_r^-\right]+\sigma_x\mathcal{P}_{nk}
	  T_k\mathcal{P}_{kr}^\dagger  \sigma_x\mathcal{P}_{rm}  \left[-i\lambda^2\Delta_n^+\Delta_r^-\right]\\
	  &+\mathcal{P}_{nk} T_k\mathcal{P}_{kr}^\dagger   \mathcal{P}_{rm}  
	  \left[\omega_n^++i\lambda^2+i\lambda^2\omega_n^+\omega_r^--\lambda^4\omega_r^-\right]\\
	  &+\mathcal{P}_{nk}  T_k\mathcal{P}_{kr}^\dagger  \sigma_x\mathcal{P}_{rm}
	  \left[-i\lambda^2\Delta_r^-\omega_n^++\lambda^4\Delta_r^-\right]\\
	  &+\mathcal{P}_{nm}  \big\{\zeta_n^\pm\left[\overline{\omega}_n + i\lambda^2(1+i\lambda^2\omega_n^-)\left(\overline{\Delta}_n\Delta_n^+ - \omega_n^+\overline{\omega}_n\right) +i\lambda^2\left(\omega_n^-\overline{\omega}_n - \overline{\Delta}_n\Delta_n^-\right)\right] \\
	  &\qquad\qquad\qquad\pm \left[ - i\lambda^2 +i\lambda^2(\Delta_n^+\Delta_n^- - \omega_n^+\omega_n^-) - \omega_n^+ + \lambda^4\omega_n^- \right]\big\}\\
	  &+\sigma_x\mathcal{P}_{nm} \left\{\zeta_n^\pm\left[\overline{\Delta}_n +\lambda^4\Delta_n^- \left(\overline{\Delta}_n\Delta_n^+ - \omega_n^+\overline{\omega}_n\right) \right]   \mp \left(\Delta_n^+ + \lambda^4\Delta_r^- \right)\right\}\Big]\, ,
  \end{aligned}
\end{equation}
where we introduced $T_n$ such that $T_{nm} = T_n \delta_{nm}$ and 
$\zeta_n^{\pm}=\tanh{\left(\frac{\omega+n\Omega}{2\Theta}\right)}\mp1)$.
It should be mentioned that, obtaining this equation, an error in the equivalent formula of \cite{Bertin2022} has been corrected.

\section{Computation of the current as a function of time}
\label{app:current}
The spin-resolved currents in the junction are defined as:
\begin{equation}
	\begin{aligned}
		\left\langle I_{\sigma}(t)\right\rangle &= e \int_{-\infty}^{\infty}\mathrm{d}t'\, \left[\sigma_z W_{\text{NS}}(t)\delta(t-t') G_{\text{SN}}(t',t) - \sigma_z G_{\text{SN}}(t,t')W_{\text{NS}}(t)\delta(t-t') \right]_{\sigma\sigma}\\
		&=\frac{e}{2\pi} \sum_{n,k,m}\int_{-\frac{\Omega}{2}}^{\frac{\Omega}{2}}\mathrm{d}\omega\,e^{-i(n-m)\Omega t}\left[\sigma_z \Sigma_{\text{NS}, nk}(\omega)G_{\text{SN},km}^{+-}(\omega) - \sigma_z G_{\text{NS},nk}^{+-}(\omega)\Sigma_{\text{SN}, km}(\omega)
		\right]_{\sigma\sigma}\, .
	\end{aligned}
\end{equation}
Using the following relations,
\begin{equation}
	G_{\text{NS}}^{\pm\mp}=-\left(G_{\text{SN}}^{\pm\mp}\right)^\dagger\qquad\text{and}\qquad\Sigma_{\text{NS}}^\dagger=\Sigma_{\text{SN}}\,,
\end{equation}
the currents can be written as
\begin{equation}
	\begin{aligned}
		\left\langle I_{\sigma}(t)\right\rangle &=\frac{e}{2\pi} \sum_{n,k,m}\int_{-\frac{\Omega}{2}}^{\frac{\Omega}{2}}\mathrm{d}\omega\,e^{-i(n-m)\Omega t}\left\{
		\left[\sigma_z \Sigma_{\text{NS}, nk}(\omega)G_{\text{SN},km}^{+-}(\omega)\right]_{\sigma\sigma} + \left[(n,m)\to(m,n)\right]^*\right\}\, ,
	\end{aligned}
\end{equation}
where the second term is the complex conjugate of the first one after taking $m\to n$ and $n\to m$.
Since the terms in $G^{+-}$, see Eq.~\eqref{eq:KeldyshGF}, that contains an odd number of $\sigma_x$ do not contribute to the diagonal matrix elements of $\sigma_z\Sigma G^{+-}$,  one is left with
\begin{equation}\label{eq:currentGF}
	\begin{aligned}
		\left\langle I_{\sigma}(t)\right\rangle &= \frac{e}{2\pi}\lambda^2\sum_{n,m,r,k,s} \int_{-\frac{\Omega}{2}}^{\frac{\Omega}{2}}\mathrm{d}\omega\,  e^{-i(n-m)\Omega t}\\
		&\Biggl\{i\xi_k^+\xi_r^-
		\sigma_z \mathcal{P}_{nk}^\dagger  
		\bigg[\sigma_x\mathcal{P}_{ks}  T_s\mathcal{P}_{sr} ^\dagger \sigma_x\mathcal{P}_{rm}
		\left[-i\lambda^2\Delta_k^+\Delta_r^-\right]
	  +\mathcal{P}_{ks} T_s\mathcal{P}_{sr}^\dagger   \mathcal{P}_{rm}
	  \left[\omega_k^++i\lambda^2+i\lambda^2\omega_k^+\omega_r^--\lambda^4\omega_r^-\right]\\
		&+\delta_{sr}\delta_{rm}\mathcal{P}_{km}  \bigg(\zeta_k^+\Big[\overline{\omega}_k +i\lambda^2\left(\omega_k^-\overline{\omega}_k - \overline{\Delta}_k\Delta_k^-\right)
			+ i\lambda^2(1+i\lambda^2\omega_k^-)\left(\overline{\Delta}_k\Delta_k^+ - \omega_k^+\overline{\omega}_k\right)\Big]\\
		+ &\Big[ - i\lambda^2 +i\lambda^2(\Delta_k^+\Delta_k^- - \omega_k^+\omega_k^-)
		- \omega_k^+ + \lambda^4\omega_k^- \Big]\bigg)\Biggr]
		+ \left[(n,m)\to(m,n)\right]^*\Bigg\}_{\sigma\sigma}\, .
	\end{aligned}
\end{equation}

\subsection{Zero gap limit.}
In this regime one has
\begin{equation}
	\omega_n^{\pm}=\pm i,\qquad\overline{\omega}_n=-i \qquad\text{and}\qquad\Delta^{\pm}=\overline{\Delta}_n=0\, ,
\end{equation}
such that the average current simply reads
\begin{equation}
		\left\langle I_{\sigma}(t)\right\rangle = \frac{e}{\pi}\frac{\lambda^2}{\left(1+\lambda^2\right)^2}\sum_{n,m,r,s,k} \int_{-\frac{\Omega}{2}}^{\frac{\Omega}{2}}\mathrm{d}\omega\, 
		e^{i(m-n)\Omega t}\bigg[\sigma_z \mathcal{P}_{nk}^\dagger  \mathcal{P}_{km}  \delta_{sr}\delta_{rm}\tanh\left(\frac{\omega+k\Omega}{2\Theta}\right)
		-\sigma_z \mathcal{P}_{nk}^\dagger  \mathcal{P}_{kn}  T_s\mathcal{P}_{sr}^\dagger  \mathcal{P}_{rm}\bigg]_{\sigma\sigma}\, .
\end{equation}
As expected, the preceding expression is diagonal in spin space, such that, after 
summing over $r$ and $s$, one gets
\begin{equation}
		\left\langle I_{\uparrow(t)}\right\rangle =  \frac{e}{\pi}\frac{\lambda^2}{\left(1+\lambda^2\right)^2}\sum_{m,n,k} \int_{-\frac{\Omega}{2}}^{\frac{\Omega}{2}}\mathrm{d}\omega\,  e^{i(m-n)\Omega t}p^*_{k-n,\uparrow}p_{k-m,\uparrow}
		\left[\tanh\left(\frac{\omega+k\Omega}{2\Theta}\right)-
				\tanh\left(\frac{\omega+m\Omega-eV_{\text{DC}_\uparrow}}{2\Theta}\right)
				\right] ,
\end{equation}
and a similar expression for $\left\langle I_{\downarrow(t)}\right\rangle$.
Changing variables as $\tilde{\omega}=\omega+k\Omega$, $\tilde{m}=m-k$ and $\tilde{n}=n-k$ and summing over $k$, the current becomes
\begin{equation}
		\left\langle I_{\uparrow}(t)\right\rangle = \frac{e}{\pi}\frac{\lambda^2}{\left(1+\lambda^2\right)^2}\sum_{m,n} \int_{-\infty}^{\infty}\mathrm{d}\omega\, e^{i(m-n)\Omega t}
		p^*_{-n,\uparrow}p_{-m,\uparrow}\left[\tanh\left(\frac{\omega}{2\Theta}\right)-
		\tanh\left(\frac{\omega-(m+q_\uparrow)\Omega}{2\Theta}\right)\right] .
\end{equation}
At zero temperature, the final integration over $\omega$ can be performed and yields
\begin{equation}\label{eq:nlimitproof}
	\left\langle I_{\sigma}(t)\right\rangle = \frac{e\Omega}{\pi}\frac{\lambda^2}{\left(1+\lambda^2\right)^2}\sum_{m,n} e^{i(n-m)\Omega t}\left(m+q_{\sigma}\right)p_{m,\sigma}p_{n,\sigma}^*,
\end{equation}
which, using the expression of the $p_{k,\sigma}$ and $q_\sigma$, simply results in
\begin{equation}\label{eq:currentnormal}
	\left\langle I_{\sigma}(t)\right\rangle = \frac{2e^2}{\pi}\frac{\lambda^2}{\left(1+\lambda^2\right)^2}
	V_{\sigma}(t)\, ,
\end{equation}
as expected.

\subsection{Infinite gap regime.}
In this regime, one has
\begin{equation}
	\omega_n^\pm=\overline{\omega}_n=\overline{\Delta}_n=0\quad\text{and}\quad\Delta_n^\pm=1\quad\text{thus}\quad\xi_n^\pm=1/(1+\lambda^4)\, ,
\end{equation}
such the currents, see Eq.~\eqref{eq:currentGF},  become
\begin{equation}
	\begin{aligned}
		\left\langle I_{\sigma}(t)\right\rangle = \frac{e}{2\pi}\frac{\lambda^4}{(1+\lambda^4)^2}\sum_{n,m,r,k,s} \int_{-\frac{\Omega}{2}}^{\frac{\Omega}{2}}\mathrm{d}\omega\, 
		\bigg\{& e^{-i(n-m)\Omega t}\left[\sigma_z \mathcal{P}_{nk}^\dagger  \sigma_x\mathcal{P}_{ks}  T_s\mathcal{P}_{sr}^\dagger  \sigma_x\mathcal{P}_{rm}   - \sigma_z \mathcal{P}_{nk}^\dagger  \mathcal{P}_{ks}  T_s \mathcal{P}_{sr}^\dagger\mathcal{P}_{rm}\right]_{\sigma\sigma}\\
		&+ \left[(n,m)\to(m,n)\right]^*\bigg\}\, .
	\end{aligned}
\end{equation}
Formally, the second term , one could use that
$\sum_{n,r,s} \mathcal{P}_{kn}^\dagger\mathcal{P}_{ns}T_s\mathcal{P}_{sr}^\dagger\mathcal{P}_{rm}=\delta_{ns}T_s\delta_{sm}$, but that would make the whole expression divergent for all $k\ne m$, i.e. for all terms but the currents averaged over a period. Therefore, when expanding the sums, one must always pay attention to keep convergent series. Performing a careful expansion, in the Andreev regime, of Eq.~\eqref{eq:fullGF}, one can show that the Keldysh dressed Green's function reads
\begin{equation}
    G_{SN}^{+-}=\frac{\lambda^3}{(1+\lambda^4)^2}\left[
    \sigma_x\mathcal{P}\left(T-\openone\right)\mathcal{P}^{\dagger}\sigma_x\mathcal{P}
    -\sigma_x\sigma_x\mathcal{P}\left(T-\openone\right)\right],
\end{equation}
such that a properly converging expression of the current is
\begin{equation}
	\begin{aligned}
		\left\langle I_{\sigma}(t)\right\rangle = \frac{e}{2\pi}\frac{\lambda^4}{(1+\lambda^4)^2}\sum_{n,m,r,k,s} \int_{-\frac{\Omega}{2}}^{\frac{\Omega}{2}}\mathrm{d}\omega\, 
		\bigg\{& e^{-i(n-m)\Omega t}\left[\sigma_z \mathcal{P}_{nk}^\dagger  \sigma_x\mathcal{P}_{ks}  T_s\mathcal{P}_{sr}^\dagger  \sigma_x\mathcal{P}_{rm}   - \sigma_z \mathcal{P}_{nk}^\dagger  \sigma_x\mathcal{P}_{ks} \mathcal{P}_{sr}^\dagger\sigma_x\mathcal{P}_{rm}T_m\right]_{\sigma\sigma}\\
		&+ \left[(n,m)\to(m,n)\right]^*\bigg\}\, .
	\end{aligned}
\end{equation}

More precisely, the $2\times 2$ matrices 
$\sigma_z \mathcal{P}_{nk}^\dagger  \sigma_x\mathcal{P}_{ks}  T_s\mathcal{P}_{sr}^\dagger\sigma_x\mathcal{P}_{rm}   - \sigma_z \mathcal{P}_{nk}^\dagger  \sigma_x\mathcal{P}_{ks} \mathcal{P}_{sr}^\dagger   \sigma_x\mathcal{P}_{rm} T_m$ are diagonal with the following entries
\begin{equation}\label{eq:temp1}
	\begin{aligned}
		 &p^*_{k-n,\uparrow}p^*_{s-k,\downarrow}p_{r-s,\downarrow}p_{r-m,\uparrow}\left[\tanh\left(\frac{\omega+s\Omega+eV_{\text{DC},\downarrow}}{2\Theta}\right) -\tanh\left(\frac{\omega+m\Omega-eV_{\text{DC},\uparrow}}{2\Theta}\right)\right]\\
		 -&p_{n-k,\downarrow}p_{k-s,\uparrow}p^*_{r-s,\uparrow}p^*_{m-r,\downarrow}\left[\tanh\left(\frac{\omega+s\Omega-eV_{\text{DC},\uparrow}}{2\Theta}\right) -\tanh\left(\frac{\omega+m\Omega+eV_{\text{DC},\downarrow}}{2\Theta}\right)\right].
	\end{aligned}
\end{equation}
Thereby, one gets
\begin{equation}
\begin{aligned}
\left\langle I_{\uparrow}(t)\right\rangle = \frac{e}{2\pi}\frac{\lambda^4}{(1+\lambda^4)^2}&\sum_{n,m,r,k,s}\int_{-\frac{\Omega}{2}}^{\frac{\Omega}{2}}\mathrm{d}\omega\, 
e^{-i(n-m)\Omega t}\times\\
&\Bigg\{p^*_{k-n,\uparrow}p^*_{s-k,\downarrow}p_{r-s,\downarrow}p_{r-m,\uparrow}\left[\tanh\left(\frac{\omega+s\Omega+eV_{\text{DC},\downarrow}}{2\Theta}\right) -\tanh\left(\frac{\omega+m\Omega-eV_{\text{DC},\uparrow}}{2\Theta}\right)\right]\\
		&+\left[(n,m)\to(m,n)\right]^*\Bigg\}\, .
\end{aligned}
\end{equation}
Perform the following change of variables, $x = \omega + (s+q_{\downarrow})\Omega$and shift all the indices by $s$ (except $s$ itself), one can perform the sum over $s$, which yields
\begin{equation}
\left\langle I_{\uparrow}(t)\right\rangle = \frac{e}{\pi}\frac{\lambda^4}{(1+\lambda^4)^2}\sum_{n,r,k,m} \int_{-\infty}^{\infty}\mathrm{d}x\, e^{i(m-n)\Omega t}
p^*_{k-n,\uparrow}p^*_{s-k,\downarrow}p_{r-s,\downarrow}p_{r-m,\uparrow}\left[
\tanh\left(\frac{x}{2\Theta}\right)-
\tanh\left(\frac{x+m\Omega-(q_{\uparrow}+q_{\downarrow})}{2\Theta}\right) \right].
\end{equation}
The integral can be performed and the preceding expression becomes
\begin{equation}
\left\langle I_{\uparrow}(t)\right\rangle = \frac{2e}{\pi}\frac{\lambda^4}{(1+\lambda^4)^2}\sum_{n,r,k,m}
e^{i(m-n)\Omega t}p^*_{k-n,\uparrow}p^*_{-k,\downarrow}p_{-r,\downarrow}p_{r-m,\uparrow}
\bigg(q_{\uparrow}+q_{\downarrow}-m\bigg),
\end{equation}
which using the definition of the $p_{l\sigma}$ leads to
\begin{equation}
	\left\langle I_{\uparrow}(t)\right\rangle = \frac{e^2}{\pi}\frac{\tau_A}{2}(V_{\uparrow}(t) + V_{\downarrow}(t))\, ,
\end{equation}
and, similarly, $\left\langle I_{\downarrow}(t)\right\rangle = \frac{e^2}{\pi}\frac{\tau_A}{2}(V_{\uparrow}(t) + V_{\downarrow}(t))$.

\section{Noise calculation}\label{app:noisecalc}
The total noise is defined as
\begin{equation}
	S_{NN}(t)=\int_{-\infty}^{+\infty}\mathrm{d}t' \left[ I_N \left(t+t' \right) I_N \left( t \right) - \left\langle I_N \left( t+t' \right) \right\rangle \left\langle I_N \left( t \right) \right\rangle\right] \, ,
\end{equation}
where $I_N$ is the total current operator across the junction.
Using Wick theorem, its average value becomes
\begin{equation}
		\begin{aligned}
			\left\langle S_{NN}(t)\right\rangle=-e^2\int_{-\infty}^{+\infty}\mathrm{d}t'\text{Tr}_{\text{N}}\Big\{ & 2\Re\left[\sigma_z W_{NS}(t) G_{SN}^{-+}(t,t')\sigma_z W_{NS}(t')G_{SN}^{+-}(t',t)\right] \\
			&- \sigma_z W_{SN} (t) G_{SS}^{-+}(t,t')\sigma_z W_{NS}(t')G_{NN}^{+-}(t',t) \\
			&-\sigma_z W_{NS}(t) G_{NN}^{-+}(t,t')\sigma_z W_{NS}(t')G_{SS}^{+-}(t',t)\Big\}\, ,
			\label{NoiseTime}
		\end{aligned}
\end{equation}
which for a periodic drive leads to
\begin{equation}
  \begin{aligned}
    \overline{\left\langle S\right\rangle}=-2e^2\int_{-\Omega/2}^{\Omega/2}\frac{d\omega}{2\pi}\text{Tr}_{\text{NH}}\big[&2\text{Re}\left(\sigma_z \Sigma_{SN} G_{NS}^{+-} \sigma_z \Sigma_{SN} G_{NS}^{-+}\right)\\
    -&\sigma_z \Sigma_{SN} G_{NN}^{+-}\sigma_z \Sigma_{NS} G_{SS}^{-+}-\sigma_z \Sigma_{NS} G_{SS}^{+-}\sigma_z \Sigma_{SN} G_{NN}^{-+}
    \big]\, ,
  \end{aligned}
\end{equation}
which, in the Andreev regime, becomes
\begin{equation}
    \overline{\left\langle S\right\rangle}=-2e^2\tau_A\int_{-\Omega/2}^{\Omega/2}\frac{d\omega}{2\pi}\text{Tr}_{\text{NH}}\left[(1-\tau_A)PTP^\dagger\sigma_xPTP^\dagger\sigma_x - \openone + \tau_AT^2\right]\, .
\end{equation}
Computations along the same lines as for the currents lead to
\begin{equation}
		\overline{\left\langle S\right\rangle}=\frac{ e^2}{\pi}
		\left[4\tau^2_A\Theta
		+2\tau_A(1-\tau_A)\sum_{krm}(p_{n,\downarrow}^*p_{r,\downarrow}p_{m-r,\uparrow}p_{m-n,\uparrow}^* )(m+q_\uparrow+q_\downarrow)\Omega
		\coth\left(\frac{\Omega(m+q_\uparrow+q_\downarrow)}{2\Theta}\right)\right]\, .
\end{equation}
 
\end{appendix}
\twocolumngrid
\bibliography{ns-shifted-levitons_prb_final} 

\begin{thebibliography}{51}%
\makeatletter
\providecommand \@ifxundefined [1]{%
 \@ifx{#1\undefined}
}%
\providecommand \@ifnum [1]{%
 \ifnum #1\expandafter \@firstoftwo
 \else \expandafter \@secondoftwo
 \fi
}%
\providecommand \@ifx [1]{%
 \ifx #1\expandafter \@firstoftwo
 \else \expandafter \@secondoftwo
 \fi
}%
\providecommand \natexlab [1]{#1}%
\providecommand \enquote  [1]{``#1''}%
\providecommand \bibnamefont  [1]{#1}%
\providecommand \bibfnamefont [1]{#1}%
\providecommand \citenamefont [1]{#1}%
\providecommand \href@noop [0]{\@secondoftwo}%
\providecommand \href [0]{\begingroup \@sanitize@url \@href}%
\providecommand \@href[1]{\@@startlink{#1}\@@href}%
\providecommand \@@href[1]{\endgroup#1\@@endlink}%
\providecommand \@sanitize@url [0]{\catcode `\\12\catcode `\$12\catcode
  `\&12\catcode `\#12\catcode `\^12\catcode `\_12\catcode `\%12\relax}%
\providecommand \@@startlink[1]{}%
\providecommand \@@endlink[0]{}%
\providecommand \url  [0]{\begingroup\@sanitize@url \@url }%
\providecommand \@url [1]{\endgroup\@href {#1}{\urlprefix }}%
\providecommand \urlprefix  [0]{URL }%
\providecommand \Eprint [0]{\href }%
\providecommand \doibase [0]{https://doi.org/}%
\providecommand \selectlanguage [0]{\@gobble}%
\providecommand \bibinfo  [0]{\@secondoftwo}%
\providecommand \bibfield  [0]{\@secondoftwo}%
\providecommand \translation [1]{[#1]}%
\providecommand \BibitemOpen [0]{}%
\providecommand \bibitemStop [0]{}%
\providecommand \bibitemNoStop [0]{.\EOS\space}%
\providecommand \EOS [0]{\spacefactor3000\relax}%
\providecommand \BibitemShut  [1]{\csname bibitem#1\endcsname}%
\let\auto@bib@innerbib\@empty
\bibitem [{\citenamefont {Bocquillon}\ \emph {et~al.}(2014)\citenamefont
  {Bocquillon}, \citenamefont {Freulon}, \citenamefont {Parmentier},
  \citenamefont {Berroir}, \citenamefont {Pla\c{c}ais}, \citenamefont {Wahl},
  \citenamefont {Rech}, \citenamefont {Jonckheere}, \citenamefont {Martin},
  \citenamefont {Grenier}, \citenamefont {Ferraro}, \citenamefont
  {Degiovanni},\ and\ \citenamefont {F\`eve}}]{bocquillon14}%
  \BibitemOpen
  \bibfield  {author} {\bibinfo {author} {\bibfnamefont {E.}~\bibnamefont
  {Bocquillon}}, \bibinfo {author} {\bibfnamefont {V.}~\bibnamefont {Freulon}},
  \bibinfo {author} {\bibfnamefont {F.~D.}\ \bibnamefont {Parmentier}},
  \bibinfo {author} {\bibfnamefont {J.}~\bibnamefont {Berroir}}, \bibinfo
  {author} {\bibfnamefont {B.}~\bibnamefont {Pla\c{c}ais}}, \bibinfo {author}
  {\bibfnamefont {C.}~\bibnamefont {Wahl}}, \bibinfo {author} {\bibfnamefont
  {J.}~\bibnamefont {Rech}}, \bibinfo {author} {\bibfnamefont {T.}~\bibnamefont
  {Jonckheere}}, \bibinfo {author} {\bibfnamefont {T.}~\bibnamefont {Martin}},
  \bibinfo {author} {\bibfnamefont {C.}~\bibnamefont {Grenier}}, \bibinfo
  {author} {\bibfnamefont {D.}~\bibnamefont {Ferraro}}, \bibinfo {author}
  {\bibfnamefont {P.}~\bibnamefont {Degiovanni}},\ and\ \bibinfo {author}
  {\bibfnamefont {G.}~\bibnamefont {F\`eve}},\ }\href
  {https://doi.org/10.1002/andp.201300181} {\bibfield  {journal} {\bibinfo
  {journal} {Ann. Phys.}\ }\textbf {\bibinfo {volume} {526}},\ \bibinfo {pages}
  {1} (\bibinfo {year} {2014})}\BibitemShut {NoStop}%
\bibitem [{\citenamefont {B\"auerle}\ \emph {et~al.}(2018)\citenamefont
  {B\"auerle}, \citenamefont {Glattli}, \citenamefont {Meunier}, \citenamefont
  {Portier}, \citenamefont {Roche}, \citenamefont {Roulleau}, \citenamefont
  {Takada},\ and\ \citenamefont {Waintal}}]{bauerle18}%
  \BibitemOpen
  \bibfield  {author} {\bibinfo {author} {\bibfnamefont {C.}~\bibnamefont
  {B\"auerle}}, \bibinfo {author} {\bibfnamefont {D.~C.}\ \bibnamefont
  {Glattli}}, \bibinfo {author} {\bibfnamefont {T.}~\bibnamefont {Meunier}},
  \bibinfo {author} {\bibfnamefont {F.}~\bibnamefont {Portier}}, \bibinfo
  {author} {\bibfnamefont {P.}~\bibnamefont {Roche}}, \bibinfo {author}
  {\bibfnamefont {P.}~\bibnamefont {Roulleau}}, \bibinfo {author}
  {\bibfnamefont {S.}~\bibnamefont {Takada}},\ and\ \bibinfo {author}
  {\bibfnamefont {X.}~\bibnamefont {Waintal}},\ }\href
  {https://doi.org/10.1088/1361-6633/aaa98a} {\bibfield  {journal} {\bibinfo
  {journal} {Rep. Progr. Phys.}\ }\textbf {\bibinfo {volume} {81}},\ \bibinfo
  {pages} {056503} (\bibinfo {year} {2018})}\BibitemShut {NoStop}%
\bibitem [{\citenamefont {Brown}\ and\ \citenamefont
  {Twiss}(1956)}]{brown1956a}%
  \BibitemOpen
  \bibfield  {author} {\bibinfo {author} {\bibfnamefont {R.~H.}\ \bibnamefont
  {Brown}}\ and\ \bibinfo {author} {\bibfnamefont {R.~Q.}\ \bibnamefont
  {Twiss}},\ }\href@noop {} {\bibfield  {journal} {\bibinfo  {journal}
  {Nature}\ }\textbf {\bibinfo {volume} {177}},\ \bibinfo {pages} {27}
  (\bibinfo {year} {1956})}\BibitemShut {NoStop}%
\bibitem [{\citenamefont {Hong}\ \emph {et~al.}(1987)\citenamefont {Hong},
  \citenamefont {Ou},\ and\ \citenamefont {Mandel}}]{hong1987a}%
  \BibitemOpen
  \bibfield  {author} {\bibinfo {author} {\bibfnamefont {C.-K.}\ \bibnamefont
  {Hong}}, \bibinfo {author} {\bibfnamefont {Z.-Y.}\ \bibnamefont {Ou}},\ and\
  \bibinfo {author} {\bibfnamefont {L.}~\bibnamefont {Mandel}},\ }\href@noop {}
  {\bibfield  {journal} {\bibinfo  {journal} {Physical review letters}\
  }\textbf {\bibinfo {volume} {59}},\ \bibinfo {pages} {2044} (\bibinfo {year}
  {1987})}\BibitemShut {NoStop}%
\bibitem [{\citenamefont {Levitov}\ \emph {et~al.}(1996)\citenamefont
  {Levitov}, \citenamefont {Lee},\ and\ \citenamefont
  {Lesovik}}]{levitov1996a}%
  \BibitemOpen
  \bibfield  {author} {\bibinfo {author} {\bibfnamefont {L.~S.}\ \bibnamefont
  {Levitov}}, \bibinfo {author} {\bibfnamefont {H.}~\bibnamefont {Lee}},\ and\
  \bibinfo {author} {\bibfnamefont {G.~B.}\ \bibnamefont {Lesovik}},\ }\href
  {https://doi.org/10.1063/1.531672} {\bibfield  {journal} {\bibinfo  {journal}
  {Journal of Mathematical Physics}\ }\textbf {\bibinfo {volume} {37}},\
  \bibinfo {pages} {4845} (\bibinfo {year} {1996})},\ \Eprint
  {https://arxiv.org/abs/9607137} {arXiv:9607137 [cond-mat]} \BibitemShut
  {NoStop}%
\bibitem [{\citenamefont {Dubois}\ \emph
  {et~al.}(2013{\natexlab{a}})\citenamefont {Dubois}, \citenamefont {Jullien},
  \citenamefont {Portier}, \citenamefont {Roche}, \citenamefont {Cavanna},
  \citenamefont {Jin}, \citenamefont {Wegscheider}, \citenamefont {Roulleau},\
  and\ \citenamefont {Glattli}}]{dubois2013b}%
  \BibitemOpen
  \bibfield  {author} {\bibinfo {author} {\bibfnamefont {J.}~\bibnamefont
  {Dubois}}, \bibinfo {author} {\bibfnamefont {T.}~\bibnamefont {Jullien}},
  \bibinfo {author} {\bibfnamefont {F.}~\bibnamefont {Portier}}, \bibinfo
  {author} {\bibfnamefont {P.}~\bibnamefont {Roche}}, \bibinfo {author}
  {\bibfnamefont {A.}~\bibnamefont {Cavanna}}, \bibinfo {author} {\bibfnamefont
  {Y.}~\bibnamefont {Jin}}, \bibinfo {author} {\bibfnamefont {W.}~\bibnamefont
  {Wegscheider}}, \bibinfo {author} {\bibfnamefont {P.}~\bibnamefont
  {Roulleau}},\ and\ \bibinfo {author} {\bibfnamefont {D.~C.}\ \bibnamefont
  {Glattli}},\ }\href {https://doi.org/10.1038/nature12713} {\bibfield
  {journal} {\bibinfo  {journal} {{Nature}}\ }\textbf {\bibinfo {volume}
  {502}},\ \bibinfo {pages} {L659 } (\bibinfo {year}
  {2013}{\natexlab{a}})}\BibitemShut {NoStop}%
\bibitem [{\citenamefont {Dubois}\ \emph
  {et~al.}(2013{\natexlab{b}})\citenamefont {Dubois}, \citenamefont {Jullien},
  \citenamefont {Grenier}, \citenamefont {Degiovanni}, \citenamefont
  {Roulleau},\ and\ \citenamefont {Glattli}}]{dubois2013a}%
  \BibitemOpen
  \bibfield  {author} {\bibinfo {author} {\bibfnamefont {J.}~\bibnamefont
  {Dubois}}, \bibinfo {author} {\bibfnamefont {T.}~\bibnamefont {Jullien}},
  \bibinfo {author} {\bibfnamefont {C.}~\bibnamefont {Grenier}}, \bibinfo
  {author} {\bibfnamefont {P.}~\bibnamefont {Degiovanni}}, \bibinfo {author}
  {\bibfnamefont {P.}~\bibnamefont {Roulleau}},\ and\ \bibinfo {author}
  {\bibfnamefont {D.~C.}\ \bibnamefont {Glattli}},\ }\href
  {https://doi.org/10.1103/PhysRevB.88.085301} {\bibfield  {journal} {\bibinfo
  {journal} {Physical Review B - Condensed Matter and Materials Physics}\
  }\textbf {\bibinfo {volume} {88}},\ \bibinfo {pages} {1} (\bibinfo {year}
  {2013}{\natexlab{b}})}\BibitemShut {NoStop}%
\bibitem [{\citenamefont {Grenier}\ \emph {et~al.}(2013)\citenamefont
  {Grenier}, \citenamefont {Dubois}, \citenamefont {Jullien}, \citenamefont
  {Roulleau}, \citenamefont {Glattli},\ and\ \citenamefont
  {Degiovanni}}]{grenier13}%
  \BibitemOpen
  \bibfield  {author} {\bibinfo {author} {\bibfnamefont {C.}~\bibnamefont
  {Grenier}}, \bibinfo {author} {\bibfnamefont {J.}~\bibnamefont {Dubois}},
  \bibinfo {author} {\bibfnamefont {T.}~\bibnamefont {Jullien}}, \bibinfo
  {author} {\bibfnamefont {P.}~\bibnamefont {Roulleau}}, \bibinfo {author}
  {\bibfnamefont {D.~C.}\ \bibnamefont {Glattli}},\ and\ \bibinfo {author}
  {\bibfnamefont {P.}~\bibnamefont {Degiovanni}},\ }\href
  {https://doi.org/10.1103/PhysRevB.88.085302} {\bibfield  {journal} {\bibinfo
  {journal} {Phys. Rev. B}\ }\textbf {\bibinfo {volume} {88}},\ \bibinfo
  {pages} {085302} (\bibinfo {year} {2013})}\BibitemShut {NoStop}%
\bibitem [{\citenamefont {Rech}\ \emph {et~al.}(2017)\citenamefont {Rech},
  \citenamefont {Ferraro}, \citenamefont {Jonckheere}, \citenamefont
  {Vannucci}, \citenamefont {Sassetti},\ and\ \citenamefont
  {Martin}}]{rech2017a}%
  \BibitemOpen
  \bibfield  {author} {\bibinfo {author} {\bibfnamefont {J.}~\bibnamefont
  {Rech}}, \bibinfo {author} {\bibfnamefont {D.}~\bibnamefont {Ferraro}},
  \bibinfo {author} {\bibfnamefont {T.}~\bibnamefont {Jonckheere}}, \bibinfo
  {author} {\bibfnamefont {L.}~\bibnamefont {Vannucci}}, \bibinfo {author}
  {\bibfnamefont {M.}~\bibnamefont {Sassetti}},\ and\ \bibinfo {author}
  {\bibfnamefont {T.}~\bibnamefont {Martin}},\ }\bibfield  {journal} {\bibinfo
  {journal} {Physical Review Letters}\ }\textbf {\bibinfo {volume} {118}},\
  \href {https://doi.org/10.1103/PhysRevLett.118.076801}
  {10.1103/PhysRevLett.118.076801} (\bibinfo {year} {2017})\BibitemShut
  {NoStop}%
\bibitem [{\citenamefont {Acciai}\ \emph {et~al.}(2019)\citenamefont {Acciai},
  \citenamefont {Ronetti}, \citenamefont {Ferraro}, \citenamefont {Rech},
  \citenamefont {Jonckheere}, \citenamefont {Sassetti},\ and\ \citenamefont
  {Martin}}]{acciai2019a}%
  \BibitemOpen
  \bibfield  {author} {\bibinfo {author} {\bibfnamefont {M.}~\bibnamefont
  {Acciai}}, \bibinfo {author} {\bibfnamefont {F.}~\bibnamefont {Ronetti}},
  \bibinfo {author} {\bibfnamefont {D.}~\bibnamefont {Ferraro}}, \bibinfo
  {author} {\bibfnamefont {J.}~\bibnamefont {Rech}}, \bibinfo {author}
  {\bibfnamefont {T.}~\bibnamefont {Jonckheere}}, \bibinfo {author}
  {\bibfnamefont {M.}~\bibnamefont {Sassetti}},\ and\ \bibinfo {author}
  {\bibfnamefont {T.}~\bibnamefont {Martin}},\ }\href@noop {} {\bibfield
  {journal} {\bibinfo  {journal} {Physical Review B}\ }\textbf {\bibinfo
  {volume} {100}},\ \bibinfo {pages} {085418} (\bibinfo {year}
  {2019})}\BibitemShut {NoStop}%
\bibitem [{\citenamefont {Ronetti}\ \emph {et~al.}(2020)\citenamefont
  {Ronetti}, \citenamefont {Carrega},\ and\ \citenamefont
  {Sassetti}}]{Ronetti2020}%
  \BibitemOpen
  \bibfield  {author} {\bibinfo {author} {\bibfnamefont {F.}~\bibnamefont
  {Ronetti}}, \bibinfo {author} {\bibfnamefont {M.}~\bibnamefont {Carrega}},\
  and\ \bibinfo {author} {\bibfnamefont {M.}~\bibnamefont {Sassetti}},\ }\href
  {https://doi.org/10.1103/PhysRevResearch.2.013203} {\bibfield  {journal}
  {\bibinfo  {journal} {Phys. Rev. Res.}\ }\textbf {\bibinfo {volume} {2}},\
  \bibinfo {pages} {013203} (\bibinfo {year} {2020})}\BibitemShut {NoStop}%
\bibitem [{\citenamefont {Bertin-Johannet}\ \emph {et~al.}(2022)\citenamefont
  {Bertin-Johannet}, \citenamefont {Rech}, \citenamefont {Jonckheere},
  \citenamefont {Gr\'emaud}, \citenamefont {Raymond},\ and\ \citenamefont
  {Martin}}]{Bertin2022}%
  \BibitemOpen
  \bibfield  {author} {\bibinfo {author} {\bibfnamefont {B.}~\bibnamefont
  {Bertin-Johannet}}, \bibinfo {author} {\bibfnamefont {J.}~\bibnamefont
  {Rech}}, \bibinfo {author} {\bibfnamefont {T.}~\bibnamefont {Jonckheere}},
  \bibinfo {author} {\bibfnamefont {B.}~\bibnamefont {Gr\'emaud}}, \bibinfo
  {author} {\bibfnamefont {L.}~\bibnamefont {Raymond}},\ and\ \bibinfo {author}
  {\bibfnamefont {T.}~\bibnamefont {Martin}},\ }\href
  {https://doi.org/10.1103/PhysRevB.105.115112} {\bibfield  {journal} {\bibinfo
   {journal} {Phys. Rev. B}\ }\textbf {\bibinfo {volume} {105}},\ \bibinfo
  {pages} {115112} (\bibinfo {year} {2022})}\BibitemShut {NoStop}%
\bibitem [{\citenamefont {Lesovik}\ and\ \citenamefont
  {Levitov}(1994)}]{lesovik94}%
  \BibitemOpen
  \bibfield  {author} {\bibinfo {author} {\bibfnamefont {G.~B.}\ \bibnamefont
  {Lesovik}}\ and\ \bibinfo {author} {\bibfnamefont {L.~S.}\ \bibnamefont
  {Levitov}},\ }\href {https://doi.org/10.1103/PhysRevLett.72.538} {\bibfield
  {journal} {\bibinfo  {journal} {Phys. Rev. Lett.}\ }\textbf {\bibinfo
  {volume} {72}},\ \bibinfo {pages} {538} (\bibinfo {year} {1994})}\BibitemShut
  {NoStop}%
\bibitem [{\citenamefont {Kouwenhoven}\ \emph {et~al.}(1994)\citenamefont
  {Kouwenhoven}, \citenamefont {Jauhar}, \citenamefont {McCormick},
  \citenamefont {Dixon}, \citenamefont {McEuen}, \citenamefont {Nazarov},
  \citenamefont {van~der Vaart},\ and\ \citenamefont {Foxon}}]{kouwenhoven94}%
  \BibitemOpen
  \bibfield  {author} {\bibinfo {author} {\bibfnamefont {L.~P.}\ \bibnamefont
  {Kouwenhoven}}, \bibinfo {author} {\bibfnamefont {S.}~\bibnamefont {Jauhar}},
  \bibinfo {author} {\bibfnamefont {K.}~\bibnamefont {McCormick}}, \bibinfo
  {author} {\bibfnamefont {D.}~\bibnamefont {Dixon}}, \bibinfo {author}
  {\bibfnamefont {P.~L.}\ \bibnamefont {McEuen}}, \bibinfo {author}
  {\bibfnamefont {Y.~V.}\ \bibnamefont {Nazarov}}, \bibinfo {author}
  {\bibfnamefont {N.~C.}\ \bibnamefont {van~der Vaart}},\ and\ \bibinfo
  {author} {\bibfnamefont {C.~T.}\ \bibnamefont {Foxon}},\ }\href
  {https://doi.org/10.1103/PhysRevB.50.2019} {\bibfield  {journal} {\bibinfo
  {journal} {Phys. Rev. B}\ }\textbf {\bibinfo {volume} {50}},\ \bibinfo
  {pages} {2019} (\bibinfo {year} {1994})}\BibitemShut {NoStop}%
\bibitem [{\citenamefont {F{\`e}ve}\ \emph {et~al.}(2007)\citenamefont
  {F{\`e}ve}, \citenamefont {Mah{\'e}}, \citenamefont {Berroir}, \citenamefont
  {Kontos}, \citenamefont {Placais}, \citenamefont {Glattli}, \citenamefont
  {Cavanna}, \citenamefont {Etienne},\ and\ \citenamefont {Jin}}]{feve2007a}%
  \BibitemOpen
  \bibfield  {author} {\bibinfo {author} {\bibfnamefont {G.}~\bibnamefont
  {F{\`e}ve}}, \bibinfo {author} {\bibfnamefont {A.}~\bibnamefont {Mah{\'e}}},
  \bibinfo {author} {\bibfnamefont {J.-M.}\ \bibnamefont {Berroir}}, \bibinfo
  {author} {\bibfnamefont {T.}~\bibnamefont {Kontos}}, \bibinfo {author}
  {\bibfnamefont {B.}~\bibnamefont {Placais}}, \bibinfo {author} {\bibfnamefont
  {D.}~\bibnamefont {Glattli}}, \bibinfo {author} {\bibfnamefont
  {A.}~\bibnamefont {Cavanna}}, \bibinfo {author} {\bibfnamefont
  {B.}~\bibnamefont {Etienne}},\ and\ \bibinfo {author} {\bibfnamefont
  {Y.}~\bibnamefont {Jin}},\ }\href@noop {} {\bibfield  {journal} {\bibinfo
  {journal} {Science}\ }\textbf {\bibinfo {volume} {316}},\ \bibinfo {pages}
  {1169} (\bibinfo {year} {2007})}\BibitemShut {NoStop}%
\bibitem [{\citenamefont {Ivanov}\ \emph {et~al.}(1997)\citenamefont {Ivanov},
  \citenamefont {Lee},\ and\ \citenamefont {Levitov}}]{levitov97}%
  \BibitemOpen
  \bibfield  {author} {\bibinfo {author} {\bibfnamefont {D.~A.}\ \bibnamefont
  {Ivanov}}, \bibinfo {author} {\bibfnamefont {H.~W.}\ \bibnamefont {Lee}},\
  and\ \bibinfo {author} {\bibfnamefont {L.~S.}\ \bibnamefont {Levitov}},\
  }\href {https://doi.org/10.1103/PhysRevB.56.6839} {\bibfield  {journal}
  {\bibinfo  {journal} {Phys. Rev. B}\ }\textbf {\bibinfo {volume} {56}},\
  \bibinfo {pages} {6839} (\bibinfo {year} {1997})}\BibitemShut {NoStop}%
\bibitem [{\citenamefont {Dubois}\ \emph
  {et~al.}(2013{\natexlab{c}})\citenamefont {Dubois}, \citenamefont {Jullien},
  \citenamefont {Grenier}, \citenamefont {Degiovanni}, \citenamefont
  {Roulleau},\ and\ \citenamefont {Glattli}}]{dubois13}%
  \BibitemOpen
  \bibfield  {author} {\bibinfo {author} {\bibfnamefont {J.}~\bibnamefont
  {Dubois}}, \bibinfo {author} {\bibfnamefont {T.}~\bibnamefont {Jullien}},
  \bibinfo {author} {\bibfnamefont {C.}~\bibnamefont {Grenier}}, \bibinfo
  {author} {\bibfnamefont {P.}~\bibnamefont {Degiovanni}}, \bibinfo {author}
  {\bibfnamefont {P.}~\bibnamefont {Roulleau}},\ and\ \bibinfo {author}
  {\bibfnamefont {D.~C.}\ \bibnamefont {Glattli}},\ }\href
  {https://doi.org/10.1103/PhysRevB.88.085301} {\bibfield  {journal} {\bibinfo
  {journal} {Phys. Rev. B}\ }\textbf {\bibinfo {volume} {88}},\ \bibinfo
  {pages} {085301} (\bibinfo {year} {2013}{\natexlab{c}})}\BibitemShut
  {NoStop}%
\bibitem [{\citenamefont {Jullien}\ \emph {et~al.}(2014)\citenamefont
  {Jullien}, \citenamefont {Roulleau}, \citenamefont {Roche}, \citenamefont
  {Cavanna}, \citenamefont {Jin},\ and\ \citenamefont {Glattli}}]{jullien14}%
  \BibitemOpen
  \bibfield  {author} {\bibinfo {author} {\bibfnamefont {T.}~\bibnamefont
  {Jullien}}, \bibinfo {author} {\bibfnamefont {P.}~\bibnamefont {Roulleau}},
  \bibinfo {author} {\bibfnamefont {B.}~\bibnamefont {Roche}}, \bibinfo
  {author} {\bibfnamefont {A.}~\bibnamefont {Cavanna}}, \bibinfo {author}
  {\bibfnamefont {Y.}~\bibnamefont {Jin}},\ and\ \bibinfo {author}
  {\bibfnamefont {D.~C.}\ \bibnamefont {Glattli}},\ }\href
  {https://doi.org/10.1038/nature13821} {\bibfield  {journal} {\bibinfo
  {journal} {Nature}\ }\textbf {\bibinfo {volume} {514}},\ \bibinfo {pages}
  {603} (\bibinfo {year} {2014})}\BibitemShut {NoStop}%
\bibitem [{\citenamefont {Glattli}\ and\ \citenamefont
  {Roulleau}(2016)}]{glattli16}%
  \BibitemOpen
  \bibfield  {author} {\bibinfo {author} {\bibfnamefont {D.}~\bibnamefont
  {Glattli}}\ and\ \bibinfo {author} {\bibfnamefont {P.}~\bibnamefont
  {Roulleau}},\ }\href
  {https://doi.org/https://doi.org/10.1016/j.physe.2015.10.034} {\bibfield
  {journal} {\bibinfo  {journal} {Physica E: Low-dimensional Systems and
  Nanostructures}\ }\textbf {\bibinfo {volume} {76}},\ \bibinfo {pages} {216 }
  (\bibinfo {year} {2016})}\BibitemShut {NoStop}%
\bibitem [{\citenamefont {Glattli}\ and\ \citenamefont
  {Roulleau}(2017)}]{glattli17}%
  \BibitemOpen
  \bibfield  {author} {\bibinfo {author} {\bibfnamefont {D.~C.}\ \bibnamefont
  {Glattli}}\ and\ \bibinfo {author} {\bibfnamefont {P.~S.}\ \bibnamefont
  {Roulleau}},\ }\href {https://doi.org/10.1002/pssb.201600650} {\bibfield
  {journal} {\bibinfo  {journal} {Phys. Status Solidi (b)}\ }\textbf {\bibinfo
  {volume} {254}},\ \bibinfo {pages} {1600650} (\bibinfo {year}
  {2017})}\BibitemShut {NoStop}%
\bibitem [{\citenamefont {Keeling}\ \emph {et~al.}(2006)\citenamefont
  {Keeling}, \citenamefont {Klich},\ and\ \citenamefont
  {Levitov}}]{keeling2006a}%
  \BibitemOpen
  \bibfield  {author} {\bibinfo {author} {\bibfnamefont {J.}~\bibnamefont
  {Keeling}}, \bibinfo {author} {\bibfnamefont {I.}~\bibnamefont {Klich}},\
  and\ \bibinfo {author} {\bibfnamefont {L.~S.}\ \bibnamefont {Levitov}},\
  }\href {https://doi.org/10.1103/PhysRevLett.97.116403} {\bibfield  {journal}
  {\bibinfo  {journal} {Physical Review Letters}\ }\textbf {\bibinfo {volume}
  {97}},\ \bibinfo {pages} {1} (\bibinfo {year} {2006})},\ \Eprint
  {https://arxiv.org/abs/0604017} {arXiv:0604017 [cond-mat]} \BibitemShut
  {NoStop}%
\bibitem [{\citenamefont {Blonder}\ \emph {et~al.}(1982)\citenamefont
  {Blonder}, \citenamefont {Tinkham},\ and\ \citenamefont
  {Klapwijk}}]{blonder1982a}%
  \BibitemOpen
  \bibfield  {author} {\bibinfo {author} {\bibfnamefont {G.~E.}\ \bibnamefont
  {Blonder}}, \bibinfo {author} {\bibfnamefont {M.}~\bibnamefont {Tinkham}},\
  and\ \bibinfo {author} {\bibfnamefont {T.~M.}\ \bibnamefont {Klapwijk}},\
  }\href {https://doi.org/10.1103/PhysRevB.25.4515} {\bibfield  {journal}
  {\bibinfo  {journal} {Physical Review B}\ }\textbf {\bibinfo {volume} {25}},\
  \bibinfo {pages} {4515} (\bibinfo {year} {1982})}\BibitemShut {NoStop}%
\bibitem [{\citenamefont {Beenakker}(2014)}]{beenakker14}%
  \BibitemOpen
  \bibfield  {author} {\bibinfo {author} {\bibfnamefont {C.~W.~J.}\
  \bibnamefont {Beenakker}},\ }\href
  {https://doi.org/10.1103/PhysRevLett.112.070604} {\bibfield  {journal}
  {\bibinfo  {journal} {Phys. Rev. Lett.}\ }\textbf {\bibinfo {volume} {112}},\
  \bibinfo {pages} {070604} (\bibinfo {year} {2014})}\BibitemShut {NoStop}%
\bibitem [{\citenamefont {Ferraro}\ \emph {et~al.}(2015)\citenamefont
  {Ferraro}, \citenamefont {Rech}, \citenamefont {Jonckheere},\ and\
  \citenamefont {Martin}}]{ferraro15}%
  \BibitemOpen
  \bibfield  {author} {\bibinfo {author} {\bibfnamefont {D.}~\bibnamefont
  {Ferraro}}, \bibinfo {author} {\bibfnamefont {J.}~\bibnamefont {Rech}},
  \bibinfo {author} {\bibfnamefont {T.}~\bibnamefont {Jonckheere}},\ and\
  \bibinfo {author} {\bibfnamefont {T.}~\bibnamefont {Martin}},\ }\href
  {https://doi.org/10.1103/PhysRevB.91.075406} {\bibfield  {journal} {\bibinfo
  {journal} {Phys. Rev. B}\ }\textbf {\bibinfo {volume} {91}},\ \bibinfo
  {pages} {075406} (\bibinfo {year} {2015})}\BibitemShut {NoStop}%
\bibitem [{\citenamefont {Andreev}(1964)}]{andreev1964a}%
  \BibitemOpen
  \bibfield  {author} {\bibinfo {author} {\bibfnamefont {A.~F.}\ \bibnamefont
  {Andreev}},\ }\href@noop {} {\bibfield  {journal} {\bibinfo  {journal} {Zh.
  Eksp. Teor. Fiz.}\ }\textbf {\bibinfo {volume} {46}},\ \bibinfo {pages}
  {1823} (\bibinfo {year} {1964})},\ \bibinfo {note} {[Sov. Phys.-JETP {\bf
  19}, 1228 (1964)]}\BibitemShut {NoStop}%
\bibitem [{\citenamefont {Belzig}\ and\ \citenamefont
  {Vanevic}(2016)}]{belzig2016a}%
  \BibitemOpen
  \bibfield  {author} {\bibinfo {author} {\bibfnamefont {W.}~\bibnamefont
  {Belzig}}\ and\ \bibinfo {author} {\bibfnamefont {M.}~\bibnamefont
  {Vanevic}},\ }\href {https://doi.org/10.1016/j.physe.2015.08.036} {\bibfield
  {journal} {\bibinfo  {journal} {Physica E: Low-Dimensional Systems and
  Nanostructures}\ }\textbf {\bibinfo {volume} {75}},\ \bibinfo {pages} {22}
  (\bibinfo {year} {2016})},\ \Eprint {https://arxiv.org/abs/1508.07039}
  {arXiv:1508.07039} \BibitemShut {NoStop}%
\bibitem [{\citenamefont {Bertin-Johannet}\ \emph {et~al.}(2023)\citenamefont
  {Bertin-Johannet}, \citenamefont {Raymond}, \citenamefont {Ronetti},
  \citenamefont {Rech}, \citenamefont {Jonckheere}, \citenamefont {Grémaud},\
  and\ \citenamefont {Martin}}]{Bertin2023}%
  \BibitemOpen
  \bibfield  {author} {\bibinfo {author} {\bibfnamefont {B.}~\bibnamefont
  {Bertin-Johannet}}, \bibinfo {author} {\bibfnamefont {L.}~\bibnamefont
  {Raymond}}, \bibinfo {author} {\bibfnamefont {F.}~\bibnamefont {Ronetti}},
  \bibinfo {author} {\bibfnamefont {J.}~\bibnamefont {Rech}}, \bibinfo {author}
  {\bibfnamefont {T.}~\bibnamefont {Jonckheere}}, \bibinfo {author}
  {\bibfnamefont {B.}~\bibnamefont {Grémaud}},\ and\ \bibinfo {author}
  {\bibfnamefont {T.}~\bibnamefont {Martin}},\ }\href
  {https://doi.org/10.1063/5.0148041} {\bibfield  {journal} {\bibinfo
  {journal} {Applied Physics Letters}\ }\textbf {\bibinfo {volume} {122}},\
  \bibinfo {pages} {202601} (\bibinfo {year} {2023})}\BibitemShut {NoStop}%
\bibitem [{\citenamefont {Torres}\ and\ \citenamefont
  {Martin}(1999)}]{torres1999}%
  \BibitemOpen
  \bibfield  {author} {\bibinfo {author} {\bibfnamefont {J.}~\bibnamefont
  {Torres}}\ and\ \bibinfo {author} {\bibfnamefont {T.}~\bibnamefont
  {Martin}},\ }\href@noop {} {\bibfield  {journal} {\bibinfo  {journal} {The
  European Physical Journal B-Condensed Matter and Complex Systems}\ }\textbf
  {\bibinfo {volume} {12}},\ \bibinfo {pages} {319} (\bibinfo {year}
  {1999})}\BibitemShut {NoStop}%
\bibitem [{\citenamefont {Deutscher}\ and\ \citenamefont
  {Feinberg}(2000)}]{deutscher2000}%
  \BibitemOpen
  \bibfield  {author} {\bibinfo {author} {\bibfnamefont {G.}~\bibnamefont
  {Deutscher}}\ and\ \bibinfo {author} {\bibfnamefont {D.}~\bibnamefont
  {Feinberg}},\ }\href@noop {} {\bibfield  {journal} {\bibinfo  {journal}
  {Applied Physics Letters}\ }\textbf {\bibinfo {volume} {76}},\ \bibinfo
  {pages} {487} (\bibinfo {year} {2000})}\BibitemShut {NoStop}%
\bibitem [{\citenamefont {Lesovik}\ \emph {et~al.}(2001)\citenamefont
  {Lesovik}, \citenamefont {Martin},\ and\ \citenamefont
  {Blatter}}]{lesovik2001}%
  \BibitemOpen
  \bibfield  {author} {\bibinfo {author} {\bibfnamefont {G.~B.}\ \bibnamefont
  {Lesovik}}, \bibinfo {author} {\bibfnamefont {T.}~\bibnamefont {Martin}},\
  and\ \bibinfo {author} {\bibfnamefont {G.}~\bibnamefont {Blatter}},\
  }\href@noop {} {\bibfield  {journal} {\bibinfo  {journal} {The European
  Physical Journal B-Condensed Matter and Complex Systems}\ }\textbf {\bibinfo
  {volume} {24}},\ \bibinfo {pages} {287} (\bibinfo {year} {2001})}\BibitemShut
  {NoStop}%
\bibitem [{\citenamefont {Bouchiat}\ \emph {et~al.}(2002)\citenamefont
  {Bouchiat}, \citenamefont {Chtchelkatchev}, \citenamefont {Feinberg},
  \citenamefont {Lesovik}, \citenamefont {Martin},\ and\ \citenamefont
  {Torres}}]{bouchiat2002}%
  \BibitemOpen
  \bibfield  {author} {\bibinfo {author} {\bibfnamefont {V.}~\bibnamefont
  {Bouchiat}}, \bibinfo {author} {\bibfnamefont {N.}~\bibnamefont
  {Chtchelkatchev}}, \bibinfo {author} {\bibfnamefont {D.}~\bibnamefont
  {Feinberg}}, \bibinfo {author} {\bibfnamefont {G.}~\bibnamefont {Lesovik}},
  \bibinfo {author} {\bibfnamefont {T.}~\bibnamefont {Martin}},\ and\ \bibinfo
  {author} {\bibfnamefont {J.}~\bibnamefont {Torres}},\ }\href@noop {}
  {\bibfield  {journal} {\bibinfo  {journal} {Nanotechnology}\ }\textbf
  {\bibinfo {volume} {14}},\ \bibinfo {pages} {77} (\bibinfo {year}
  {2002})}\BibitemShut {NoStop}%
\bibitem [{\citenamefont {S\'anchez}\ \emph {et~al.}(2003)\citenamefont
  {S\'anchez}, \citenamefont {L\'opez}, \citenamefont {Samuelsson},\ and\
  \citenamefont {B\"uttiker}}]{sanchez03}%
  \BibitemOpen
  \bibfield  {author} {\bibinfo {author} {\bibfnamefont {D.}~\bibnamefont
  {S\'anchez}}, \bibinfo {author} {\bibfnamefont {R.}~\bibnamefont {L\'opez}},
  \bibinfo {author} {\bibfnamefont {P.}~\bibnamefont {Samuelsson}},\ and\
  \bibinfo {author} {\bibfnamefont {M.}~\bibnamefont {B\"uttiker}},\ }\href
  {https://doi.org/10.1103/PhysRevB.68.214501} {\bibfield  {journal} {\bibinfo
  {journal} {Phys. Rev. B}\ }\textbf {\bibinfo {volume} {68}},\ \bibinfo
  {pages} {214501} (\bibinfo {year} {2003})}\BibitemShut {NoStop}%
\bibitem [{\citenamefont {Beckmann}\ \emph {et~al.}(2004)\citenamefont
  {Beckmann}, \citenamefont {Weber},\ and\ \citenamefont
  {v.~L\"ohneysen}}]{beckmann2004}%
  \BibitemOpen
  \bibfield  {author} {\bibinfo {author} {\bibfnamefont {D.}~\bibnamefont
  {Beckmann}}, \bibinfo {author} {\bibfnamefont {H.~B.}\ \bibnamefont
  {Weber}},\ and\ \bibinfo {author} {\bibfnamefont {H.}~\bibnamefont
  {v.~L\"ohneysen}},\ }\href {https://doi.org/10.1103/PhysRevLett.93.197003}
  {\bibfield  {journal} {\bibinfo  {journal} {Phys. Rev. Lett.}\ }\textbf
  {\bibinfo {volume} {93}},\ \bibinfo {pages} {197003} (\bibinfo {year}
  {2004})}\BibitemShut {NoStop}%
\bibitem [{\citenamefont {Herrmann}\ \emph {et~al.}(2010)\citenamefont
  {Herrmann}, \citenamefont {Portier}, \citenamefont {Roche}, \citenamefont
  {Yeyati}, \citenamefont {Kontos},\ and\ \citenamefont {Strunk}}]{herrmann10}%
  \BibitemOpen
  \bibfield  {author} {\bibinfo {author} {\bibfnamefont {L.~G.}\ \bibnamefont
  {Herrmann}}, \bibinfo {author} {\bibfnamefont {F.}~\bibnamefont {Portier}},
  \bibinfo {author} {\bibfnamefont {P.}~\bibnamefont {Roche}}, \bibinfo
  {author} {\bibfnamefont {A.~L.}\ \bibnamefont {Yeyati}}, \bibinfo {author}
  {\bibfnamefont {T.}~\bibnamefont {Kontos}},\ and\ \bibinfo {author}
  {\bibfnamefont {C.}~\bibnamefont {Strunk}},\ }\href
  {https://doi.org/10.1103/PhysRevLett.104.026801} {\bibfield  {journal}
  {\bibinfo  {journal} {Phys. Rev. Lett.}\ }\textbf {\bibinfo {volume} {104}},\
  \bibinfo {pages} {026801} (\bibinfo {year} {2010})}\BibitemShut {NoStop}%
\bibitem [{\citenamefont {Wei}\ and\ \citenamefont
  {Chandrasekhar}(2010)}]{wei10}%
  \BibitemOpen
  \bibfield  {author} {\bibinfo {author} {\bibfnamefont {J.}~\bibnamefont
  {Wei}}\ and\ \bibinfo {author} {\bibfnamefont {V.}~\bibnamefont
  {Chandrasekhar}},\ }\href {https://doi.org/10.1038/nphys1669} {\bibfield
  {journal} {\bibinfo  {journal} {Nature Physics}\ }\textbf {\bibinfo {volume}
  {6}},\ \bibinfo {pages} {494} (\bibinfo {year} {2010})}\BibitemShut {NoStop}%
\bibitem [{\citenamefont {Das}\ \emph {et~al.}(2012)\citenamefont {Das},
  \citenamefont {Ronen}, \citenamefont {Heiblum}, \citenamefont {Mahalu},
  \citenamefont {Kretinin},\ and\ \citenamefont {Shtrikman}}]{das12}%
  \BibitemOpen
  \bibfield  {author} {\bibinfo {author} {\bibfnamefont {A.}~\bibnamefont
  {Das}}, \bibinfo {author} {\bibfnamefont {Y.}~\bibnamefont {Ronen}}, \bibinfo
  {author} {\bibfnamefont {M.}~\bibnamefont {Heiblum}}, \bibinfo {author}
  {\bibfnamefont {D.}~\bibnamefont {Mahalu}}, \bibinfo {author} {\bibfnamefont
  {A.~V.}\ \bibnamefont {Kretinin}},\ and\ \bibinfo {author} {\bibfnamefont
  {H.}~\bibnamefont {Shtrikman}},\ }\href {https://doi.org/10.1038/ncomms2169}
  {\bibfield  {journal} {\bibinfo  {journal} {Nature Communications}\ }\textbf
  {\bibinfo {volume} {3}},\ \bibinfo {pages} {1165} (\bibinfo {year}
  {2012})}\BibitemShut {NoStop}%
\bibitem [{\citenamefont {Baba}\ \emph {et~al.}(2018)\citenamefont {Baba},
  \citenamefont {Jünger}, \citenamefont {Matsuo}, \citenamefont {Baumgartner},
  \citenamefont {Sato}, \citenamefont {Kamata}, \citenamefont {Li},
  \citenamefont {Jeppesen}, \citenamefont {Samuelson}, \citenamefont {Xu},
  \citenamefont {Schönenberger},\ and\ \citenamefont {Tarucha}}]{baba18}%
  \BibitemOpen
  \bibfield  {author} {\bibinfo {author} {\bibfnamefont {S.}~\bibnamefont
  {Baba}}, \bibinfo {author} {\bibfnamefont {C.}~\bibnamefont {Jünger}},
  \bibinfo {author} {\bibfnamefont {S.}~\bibnamefont {Matsuo}}, \bibinfo
  {author} {\bibfnamefont {A.}~\bibnamefont {Baumgartner}}, \bibinfo {author}
  {\bibfnamefont {Y.}~\bibnamefont {Sato}}, \bibinfo {author} {\bibfnamefont
  {H.}~\bibnamefont {Kamata}}, \bibinfo {author} {\bibfnamefont
  {K.}~\bibnamefont {Li}}, \bibinfo {author} {\bibfnamefont {S.}~\bibnamefont
  {Jeppesen}}, \bibinfo {author} {\bibfnamefont {L.}~\bibnamefont {Samuelson}},
  \bibinfo {author} {\bibfnamefont {H.~Q.}\ \bibnamefont {Xu}}, \bibinfo
  {author} {\bibfnamefont {C.}~\bibnamefont {Schönenberger}},\ and\ \bibinfo
  {author} {\bibfnamefont {S.}~\bibnamefont {Tarucha}},\ }\href
  {https://doi.org/10.1088/1367-2630/aac74e} {\bibfield  {journal} {\bibinfo
  {journal} {New Journal of Physics}\ }\textbf {\bibinfo {volume} {20}},\
  \bibinfo {pages} {063021} (\bibinfo {year} {2018})}\BibitemShut {NoStop}%
\bibitem [{\citenamefont {Morten}\ \emph {et~al.}(2006)\citenamefont {Morten},
  \citenamefont {Brataas},\ and\ \citenamefont {Belzig}}]{morten06}%
  \BibitemOpen
  \bibfield  {author} {\bibinfo {author} {\bibfnamefont {J.~P.}\ \bibnamefont
  {Morten}}, \bibinfo {author} {\bibfnamefont {A.}~\bibnamefont {Brataas}},\
  and\ \bibinfo {author} {\bibfnamefont {W.}~\bibnamefont {Belzig}},\ }\href
  {https://doi.org/10.1103/PhysRevB.74.214510} {\bibfield  {journal} {\bibinfo
  {journal} {Phys. Rev. B}\ }\textbf {\bibinfo {volume} {74}},\ \bibinfo
  {pages} {214510} (\bibinfo {year} {2006})}\BibitemShut {NoStop}%
\bibitem [{\citenamefont {Morten}\ \emph {et~al.}(2008)\citenamefont {Morten},
  \citenamefont {Huertas-Hernando}, \citenamefont {Belzig},\ and\ \citenamefont
  {Brataas}}]{morten08}%
  \BibitemOpen
  \bibfield  {author} {\bibinfo {author} {\bibfnamefont {J.~P.}\ \bibnamefont
  {Morten}}, \bibinfo {author} {\bibfnamefont {D.}~\bibnamefont
  {Huertas-Hernando}}, \bibinfo {author} {\bibfnamefont {W.}~\bibnamefont
  {Belzig}},\ and\ \bibinfo {author} {\bibfnamefont {A.}~\bibnamefont
  {Brataas}},\ }\href {https://doi.org/10.1103/PhysRevB.78.224515} {\bibfield
  {journal} {\bibinfo  {journal} {Phys. Rev. B}\ }\textbf {\bibinfo {volume}
  {78}},\ \bibinfo {pages} {224515} (\bibinfo {year} {2008})}\BibitemShut
  {NoStop}%
\bibitem [{\citenamefont {Cuevas}(1999)}]{cuevas1999a}%
  \BibitemOpen
  \bibfield  {author} {\bibinfo {author} {\bibfnamefont {J.~C.}\ \bibnamefont
  {Cuevas}},\ }\href@noop {} {Ph.D. thesis},\ \bibinfo  {school} {Universidad
  Autonoma} (\bibinfo {year} {1999})\BibitemShut {NoStop}%
\bibitem [{\citenamefont {Keldysh}(1965)}]{keldysh1965a}%
  \BibitemOpen
  \bibfield  {author} {\bibinfo {author} {\bibfnamefont {L.~V.}\ \bibnamefont
  {Keldysh}},\ }\href@noop {} {\bibfield  {journal} {\bibinfo  {journal} {Zh.
  Eksp. Teor. Fiz.}\ }\textbf {\bibinfo {volume} {47}},\ \bibinfo {pages}
  {1515} (\bibinfo {year} {1965})},\ \bibinfo {note} {[Sov. Phys.-JETP {\bf
  20}, 1018 (1965)]}\BibitemShut {NoStop}%
\bibitem [{\citenamefont {Rammer}\ and\ \citenamefont
  {Smith}(1986)}]{rammer86}%
  \BibitemOpen
  \bibfield  {author} {\bibinfo {author} {\bibfnamefont {J.}~\bibnamefont
  {Rammer}}\ and\ \bibinfo {author} {\bibfnamefont {H.}~\bibnamefont {Smith}},\
  }\href {https://doi.org/10.1103/RevModPhys.58.323} {\bibfield  {journal}
  {\bibinfo  {journal} {Rev. Mod. Phys.}\ }\textbf {\bibinfo {volume} {58}},\
  \bibinfo {pages} {323} (\bibinfo {year} {1986})}\BibitemShut {NoStop}%
\bibitem [{\citenamefont {Kamenev}(2011)}]{kamenev11}%
  \BibitemOpen
  \bibfield  {author} {\bibinfo {author} {\bibfnamefont {A.}~\bibnamefont
  {Kamenev}},\ }\href {https://doi.org/10.1017/CBO9781139003667} {\emph
  {\bibinfo {title} {Field Theory of Non-Equilibrium Systems}}}\ (\bibinfo
  {publisher} {Cambridge University Press},\ \bibinfo {year}
  {2011})\BibitemShut {NoStop}%
\bibitem [{\citenamefont {Tien}\ and\ \citenamefont
  {Gordon}(1963)}]{tien1963a}%
  \BibitemOpen
  \bibfield  {author} {\bibinfo {author} {\bibfnamefont {P.~K.}\ \bibnamefont
  {Tien}}\ and\ \bibinfo {author} {\bibfnamefont {J.~P.}\ \bibnamefont
  {Gordon}},\ }\href@noop {} {\bibfield  {journal} {\bibinfo  {journal} {Phys.
  Rev.}\ }\textbf {\bibinfo {volume} {129}},\ \bibinfo {pages} {647} (\bibinfo
  {year} {1963})}\BibitemShut {NoStop}%
\bibitem [{\citenamefont {Shirley}(1965)}]{shirley1965a}%
  \BibitemOpen
  \bibfield  {author} {\bibinfo {author} {\bibfnamefont {J.~H.}\ \bibnamefont
  {Shirley}},\ }\bibfield  {journal} {\bibinfo  {journal} {Physical Review}\
  }\textbf {\bibinfo {volume} {138}},\ \href
  {https://doi.org/10.1103/PhysRev.138.B979} {10.1103/PhysRev.138.B979}
  (\bibinfo {year} {1965})\BibitemShut {NoStop}%
\bibitem [{\citenamefont {Pedersen}\ and\ \citenamefont
  {Buttiker}(1998)}]{pedersen1998a}%
  \BibitemOpen
  \bibfield  {author} {\bibinfo {author} {\bibfnamefont {M.~H.}\ \bibnamefont
  {Pedersen}}\ and\ \bibinfo {author} {\bibfnamefont {M.}~\bibnamefont
  {Buttiker}},\ }\href {https://doi.org/10.1103/PhysRevB.58.12993} {\bibfield
  {journal} {\bibinfo  {journal} {Physical Review B - Condensed Matter and
  Materials Physics}\ }\textbf {\bibinfo {volume} {58}},\ \bibinfo {pages}
  {12993} (\bibinfo {year} {1998})},\ \Eprint {https://arxiv.org/abs/9803306}
  {arXiv:9803306 [cond-mat]} \BibitemShut {NoStop}%
\bibitem [{\citenamefont {Moskalets}\ and\ \citenamefont
  {B\"uttiker}(2002)}]{moskalets2002}%
  \BibitemOpen
  \bibfield  {author} {\bibinfo {author} {\bibfnamefont {M.}~\bibnamefont
  {Moskalets}}\ and\ \bibinfo {author} {\bibfnamefont {M.}~\bibnamefont
  {B\"uttiker}},\ }\href {https://doi.org/10.1103/PhysRevB.66.205320}
  {\bibfield  {journal} {\bibinfo  {journal} {Phys. Rev. B}\ }\textbf {\bibinfo
  {volume} {66}},\ \bibinfo {pages} {205320} (\bibinfo {year}
  {2002})}\BibitemShut {NoStop}%
\bibitem [{\citenamefont {Daley}\ \emph {et~al.}(2008)\citenamefont {Daley},
  \citenamefont {Zoller},\ and\ \citenamefont
  {Trauzettel}}]{PhysRevLett.100.110404}%
  \BibitemOpen
  \bibfield  {author} {\bibinfo {author} {\bibfnamefont {A.~J.}\ \bibnamefont
  {Daley}}, \bibinfo {author} {\bibfnamefont {P.}~\bibnamefont {Zoller}},\ and\
  \bibinfo {author} {\bibfnamefont {B.}~\bibnamefont {Trauzettel}},\ }\href
  {https://doi.org/10.1103/PhysRevLett.100.110404} {\bibfield  {journal}
  {\bibinfo  {journal} {Phys. Rev. Lett.}\ }\textbf {\bibinfo {volume} {100}},\
  \bibinfo {pages} {110404} (\bibinfo {year} {2008})}\BibitemShut {NoStop}%
\bibitem [{\citenamefont {Zapata}\ and\ \citenamefont
  {Sols}(2009)}]{PhysRevLett.102.180405}%
  \BibitemOpen
  \bibfield  {author} {\bibinfo {author} {\bibfnamefont {I.}~\bibnamefont
  {Zapata}}\ and\ \bibinfo {author} {\bibfnamefont {F.}~\bibnamefont {Sols}},\
  }\href {https://doi.org/10.1103/PhysRevLett.102.180405} {\bibfield  {journal}
  {\bibinfo  {journal} {Phys. Rev. Lett.}\ }\textbf {\bibinfo {volume} {102}},\
  \bibinfo {pages} {180405} (\bibinfo {year} {2009})}\BibitemShut {NoStop}%
\bibitem [{\citenamefont {Bertin-Johannet}(2024)}]{BrunoTS}%
  \BibitemOpen
  \bibfield  {author} {\bibinfo {author} {\bibfnamefont {B.}~\bibnamefont
  {Bertin-Johannet}},\ }\href@noop {} {\bibfield  {journal} {\bibinfo
  {journal} {Private communications}\ } (\bibinfo {year} {2024})}\BibitemShut
  {NoStop}%
\bibitem [{\citenamefont {Prada}\ \emph {et~al.}(2020)\citenamefont {Prada},
  \citenamefont {San-Jose}, \citenamefont {de~Moor}, \citenamefont {Geresdi},
  \citenamefont {Lee}, \citenamefont {Klinovaja}, \citenamefont {Loss},
  \citenamefont {Nygård}, \citenamefont {Aguado},\ and\ \citenamefont
  {Kouwenhoven}}]{Prada20}%
  \BibitemOpen
  \bibfield  {author} {\bibinfo {author} {\bibfnamefont {E.}~\bibnamefont
  {Prada}}, \bibinfo {author} {\bibfnamefont {P.}~\bibnamefont {San-Jose}},
  \bibinfo {author} {\bibfnamefont {M.}~\bibnamefont {de~Moor}}, \bibinfo
  {author} {\bibfnamefont {A.}~\bibnamefont {Geresdi}}, \bibinfo {author}
  {\bibfnamefont {E.}~\bibnamefont {Lee}}, \bibinfo {author} {\bibfnamefont
  {J.}~\bibnamefont {Klinovaja}}, \bibinfo {author} {\bibfnamefont
  {D.}~\bibnamefont {Loss}}, \bibinfo {author} {\bibfnamefont {J.}~\bibnamefont
  {Nygård}}, \bibinfo {author} {\bibfnamefont {R.}~\bibnamefont {Aguado}},\
  and\ \bibinfo {author} {\bibfnamefont {L.}~\bibnamefont {Kouwenhoven}},\
  }\href {https://doi.org/10.1038/s42254-020-0228-y} {\bibfield  {journal}
  {\bibinfo  {journal} {Nature Reviews Physics}\ }\textbf {\bibinfo {volume}
  {2}},\ \bibinfo {pages} {575} (\bibinfo {year} {2020})}\BibitemShut {NoStop}%
\end{thebibliography}%

\end{document}